\documentclass{pasj01}
\newcommand{\cii}{[C\,\emissiontype{II}] }
\usepackage{multirow}
\usepackage{natbib}
\Received{$\langle$15-Apr-2019$\rangle$}
\Accepted{$\langle$04-Aug-2019$\rangle$}
\Published{$\langle$publication date$\rangle$}

\begin{document}

\title{Subaru High-z Exploration of Low-Luminosity Quasars (SHELLQs) VIII. 
A less biased view of the early co-evolution of black holes and host galaxies}
\author{Takuma \textsc{Izumi},\altaffilmark{1,2,}$^{\dag,*}$ 
Masafusa \textsc{Onoue},\altaffilmark{3} 
Yoshiki \textsc{Matsuoka},\altaffilmark{4} 
Tohru \textsc{Nagao},\altaffilmark{4} 
Michael A. \textsc{Strauss}, \altaffilmark{5} 
Masatoshi \textsc{Imanishi},\altaffilmark{1,2} 
Nobunari \textsc{Kashikawa},\altaffilmark{6} 
Seiji \textsc{Fujimoto},\altaffilmark{7} 
Kotaro \textsc{Kohno},\altaffilmark{8,9} 
Yoshiki \textsc{Toba},\altaffilmark{10,11,4} 
Hideki \textsc{Umehata},\altaffilmark{12,8}
Tomotsugu \textsc{Goto},\altaffilmark{13} 
Yoshihiro \textsc{Ueda}, \altaffilmark{10} 
Hikari \textsc{Shirakata},\altaffilmark{14} 
John D. \textsc{Silverman},\altaffilmark{15}
Jenny E. \textsc{Greene},\altaffilmark{16} 
Yuichi \textsc{Harikane},\altaffilmark{7} 
Yasuhiro \textsc{Hashimoto},\altaffilmark{17} 
Soh \textsc{Ikarashi},\altaffilmark{18} 
Daisuke \textsc{Iono},\altaffilmark{1,2} 
Kazushi \textsc{Iwasawa},\altaffilmark{19} 
Chien-Hsiu \textsc{Lee},\altaffilmark{1} 
Takeo \textsc{Minezaki},\altaffilmark{8} 
Kouichiro \textsc{Nakanishi},\altaffilmark{1,2} 
Yoichi \textsc{Tamura},\altaffilmark{20} 
Ji-Jia \textsc{Tang}\altaffilmark{21}
and Akio \textsc{Taniguchi}\altaffilmark{20} 
}
\altaffiltext{1}{National Astronomical Observatory of Japan, 2-21-1 Osawa, Mitaka, Tokyo 181-8588, Japan}
\altaffiltext{2}{Department of Astronomical Science, Graduate University for Advanced Studies (SOKENDAI), 2-21-1 Osawa, Mitaka, Tokyo 181-8588, Japan}
\altaffiltext{3}{Max Planck Institut f\"ur Astronomie, K\"onigstuhl 17, D-69117 Heidelberg, Germany}
\altaffiltext{4}{Research Center for Space and Cosmic Evolution, Ehime University, Matsuyama, Ehime 790-8577, Japan} 
\altaffiltext{5}{Princeton University Observatory, Peyton Hall, Princeton, NJ 08544, USA} 
\altaffiltext{6}{Department of Astronomy, School of Science, The University of Tokyo, Tokyo 113-0033, Japan}
\altaffiltext{7}{Institute for Cosmic Ray Research, The University of Tokyo, Kashiwa, Chiba 277-8582, Japan}
\altaffiltext{8}{Institute of Astronomy, Graduate School of Science, The University of Tokyo, 2-21-1 Osawa, Mitaka, Tokyo 181-0015, Japan} 
\altaffiltext{9}{Research Center for the Early Universe, Graduate School of Science, The University of Tokyo, 7-3-1 Hongo, Bunkyo, Tokyo 113-0033, Japan} 
\altaffiltext{10}{Department of Astronomy, Kyoto University, Kitashirakawa-Oiwake-cho, Sakyo-ku, Kyoto 606-8502, Japan} 
\altaffiltext{11}{Academia Sinica Institute of Astronomy and Astrophysics, 11F of Astronomy-Mathematics Building, AS/NTU, No.1, Section 4, Roosevelt Road, Taipei 10617, Taiwan} 
\altaffiltext{12}{RIKEN Cluster for Pioneering Research, 2-1 Hirosawa, Wako, Saitama 351-0198, Japan} 
\altaffiltext{13}{Institute of Astronomy and Department of Physics, National Tsing Hua University, Hsinchu 30013, Taiwan}
\altaffiltext{14}{Department of Cosmosciences, Graduate School of Science, Hokkaido University, N10 W8, Kitaku, Sapporo, 060-0810, Japan} 
\altaffiltext{15}{Kavli Institute for the Physics and Mathematics of the Universe (Kavli-IPMU, WPI), The University of Tokyo Institutes for Advanced Study,
The University of Tokyo, Kashiwa, Chiba 277-8583, Japan}
\altaffiltext{16}{Department of Astrophysics, Princeton University, Princeton, NJ, USA} 
\altaffiltext{17}{Department of Earth Sciences, National Taiwan Normal University, Taipei 11677, Taiwan}
\altaffiltext{18}{Kapteyn Astronomical Institute, University of Groningen, P.O. Box 800, 9700 AV Groningen, Netherlands}
\altaffiltext{19}{ICREA and Institut de Ci\`{e}ncies del Cosmos, Universitat de Barcelona, IEEC-UB, Mart\'{i} i Franqu\`{e}s, 1, E-08028 Barcelona, Spain}
\altaffiltext{20}{Division of Particle and Astrophysical Science, Graduate School of Science, Nagoya University, Chikusa-ku, Nagoya, Aichi 464-8602, Japan}
\altaffiltext{21}{Research School of Astronomy and Astrophysics, Australian National University, Cotter Road, Weston Creek, ACT 2611, Australia} 
\altaffiltext{$\dag$}{NAOJ Fellow}
\email{takuma.izumi@nao.ac.jp}

\maketitle

\begin{abstract}
We present ALMA \cii line and far-infrared (FIR) continuum observations of 
three $z > 6$ low-luminosity quasars ($M_{\rm 1450} > -25$ mag) 
discovered by our Subaru Hyper Suprime-Cam (HSC) survey. 
The \cii line was detected in all three targets with luminosities of $(2.4 - 9.5) \times 10^8~L_\odot$, 
about one order of magnitude smaller than optically luminous ($M_{\rm 1450} \lesssim -25$ mag) quasars. 
The FIR continuum luminosities range from $< 9 \times 10^{10}~L_\odot$ (3$\sigma$ limit) to $\sim 2 \times 10^{12}~L_\odot$, 
indicating a wide range in star formation rates in these galaxies. 
Most of the HSC quasars studied thus far show [C\,\emissiontype{II}]/FIR luminosity ratios similar to local star-forming galaxies. 
Using the [C\,\emissiontype{II}]-based dynamical mass ($M_{\rm dyn}$) as a surrogate for bulge stellar mass ($M_{\rm bulge}$), 
we find that a significant fraction of low-luminosity quasars are located on or even below the local $M_{\rm BH} - M_{\rm bulge}$ relation, 
particularly at the massive end of the galaxy mass distribution. 
In contrast, previous studies of optically luminous quasars have found that black holes are overmassive relative to the local relation. 
Given the low luminosities of our targets, we are exploring 
the nature of the early co-evolution of supermassive black holes and their hosts in a less biased way. 
Almost all of the quasars presented in this work are growing their black hole mass at much higher pace at $z \sim 6$ 
than the parallel growth model, in which supermassive black holes and their hosts 
grow simultaneously to match the local $M_{\rm BH} - M_{\rm bulge}$ relation at all redshifts. 
As the low-luminosity quasars appear to realize the local co-evolutionary relation even at $z \sim 6$, 
they should have experienced vigorous starbursts prior to the currently observed quasar phase to catch up with the relation. 
\end{abstract}

\section{Introduction}\label{sec1}
The discovery of a tight correlation between the masses of central supermassive black holes ($M_{\rm BH}$) 
and those of galactic bulges ($M_{\rm bulge}$) or the stellar velocity dispersion in the local universe 
\citep[e.g.,][]{1998AJ....115.2285M,2000ApJ...539L...9F,2003ApJ...589L..21M,2013ARA&A..51..511K} 
strongly suggests that the formation and growth of the supermassive black holes (SMBHs) 
and their host galaxies are intimately linked, and the two undergo a {\it co-evolution}. 
Although the detailed mechanism by which the correlation arises unclear, 
theoretical models suggest that radiative and kinetic feedback of active galactic nuclei (AGNs) 
connected to the merger histories of galaxies play a pivotal role 
\citep[e.g.,][]{2004ApJ...600..580G,2005Natur.433..604D,2006ApJS..163....1H,2007ApJ...665..187L}. 
A recent high resolution simulation work based on this scheme suggests that even a quasar at $z = 7$ 
would follow the local co-evolution relation once we properly assess the mass of the host galaxy \citep{2019arXiv190102464L}. 
Detections of galaxy-scale massive AGN-driven outflows 
\citep[e.g.,][]{2008A&A...491..407N,2012MNRAS.425L..66M,2014A&A...562A..21C,2016A&A...591A..28C,2017ApJ...850..140T}, 
as well as the remarkable similarity of global star formation 
and SMBH accretion histories \citep[][for a review]{2014ARA&A..52..415M} would support this evolutionary scheme. 
As theoretical models usually make specific predictions for the time evolution of the systems, 
observations of the physical properties of both SMBHs and their host galaxies 
over cosmic time are essential to test and/or refine our current understanding of their build-up 
\citep{2017PASA...34...22G,2017PASA...34...31V}. 

From this perspective, high redshift quasars are a unique beacon 
of the early formation of SMBHs and their host galaxies, 
even in the first billion years of the universe \citep[e.g.,][]{2011Natur.474..616M,2018Natur.553..473B,2019ApJ...872L...2M}. 
The last two decades have witnessed the discovery of $> 200$ quasars at $z > 5.7$ 
owing to wide-field optical and near-infrared (NIR) surveys, including 
the Sloan Digital Sky Survey \citep[SDSS, e.g.,][]{2003AJ....125.1649F,2006AJ....131.1203F,2016ApJ...833..222J}, 
the Canada-France High-z Quasar Survey \citep[CFHQS, e.g.,][]{2007AJ....134.2435W,2010AJ....139..906W}, 
the Visible and Infrared Survey Telescope for Astronomy (VISTA) Kilo-degree Infrared Galaxy \citep[VIKING,][]{2013ApJ...779...24V,2015MNRAS.453.2259V}, 
the United Kingdom Infrared Telescope (UKIRT) Infrared Deep Sky Survey \citep[UKIDSS,][]{2009A&A...505...97M,2011Natur.474..616M,2018Natur.553..473B}, 
the Panoramic Survey Telescope \& Rapid Response System \citep[Pan-STARRS1,][]{2014AJ....148...14B,2016ApJS..227...11B,2017ApJ...849...91M}, 
and several other projects \citep[e.g.,][]{2015ApJ...798...28K,2015MNRAS.451L..16C,2015ApJ...813L..35K,2017MNRAS.466.4568T,2017MNRAS.468.4702R}. 
These surveys have found that luminous (absolute UV magnitude $M_{\rm 1450} \lesssim -25$ mag) 
quasars at $z \gtrsim 6$ are typically powered by SMBHs heavier than one billion solar masses 
and appear metal-enriched \citep[e.g.,][]{2011Natur.474..616M,2014ApJ...790..145D,2015Natur.518..512W,2019ApJ...873...35S}. 

As host galaxies of $z > 4$ quasars are hard to detect at rest-frame ultraviolet (UV) to optical wavelengths \citep{2012ApJ...756L..38M}, 
longer wavelengths (i.e., far-infrared = FIR and sub/millimeter = sub/mm) cold gas and dust emission 
from star-forming regions have been used to probe such galaxies instead. 
These host galaxies possess copious amount of 
dust ($\sim 10^8~M_\odot$) and gas ($\sim 10^{10}~M_\odot$) 
with high FIR luminosities likely due to intense starburst 
(star formation rate $\gtrsim 100 - 1000$ $M_\odot$ yr$^{-1}$) 
at $z > 6$ \citep[e.g.,][]{2003A&A...406L..55B,2003A&A...409L..47B,
2003AJ....126...15P,2003MNRAS.344L..74P,2008MNRAS.383..289P,2004ApJ...615L..17W,
2007AJ....134..617W,2008ApJ...687..848W,2011AJ....142..101W,2011ApJ...739L..34W}. 
These gaseous and dusty starburst regions appear to be spatially compact, 
with sizes of a few kpc or less \citep[e.g.,][]{2013ApJ...773...44W,2017ApJ...837..146V}, 
corresponding to the typical size of nearby bulges. 

Among various emission lines of atoms and molecules 
in the cold interstellar medium (ISM), the fine structure line of singly ionized carbon, 
the 157.74 $\mu$m \cii ${}^{2}P_{3/2} \rightarrow {}^{2}P_{1/2}$ emission line (rest frequency $\nu_{\rm rest} = 1900.5369$ GHz), 
of $z \gtrsim 6$ objects can be conveniently observed with ground-based sub/mm telescopes owing to an atmospheric window of $\sim 250$ GHz. 
The \cii line is the main coolant of the cold ISM, particularly of photodissociation regions \citep{1999RvMP...71..173H}, 
which makes this line an important tracer of star-forming activity. 
Sub/mm interferometers such as the Atacama Large Millimeter/submillimeter Array (ALMA) 
and the IRAM Plateau de Bure Interferometer (PdBI, now NOEMA) have sufficient 
resolution and sensitivity to resolve the gas dynamics of galaxies hosting not only 
optically luminous quasars \citep[e.g.,][]{2013ApJ...773...44W,2016ApJ...830...53W,
2015ApJ...805L...8B,2015A&A...574A..14C,2016ApJ...816...37V,2017ApJ...837..146V,2017ApJ...845..154V,
2017Natur.545..457D,2018ApJ...854...97D,2017ApJ...849...91M,2017ApJ...845..138S} 
but also low-luminosity ($M_{\rm 1450} \gtrsim -25$ mag) quasars 
\citep{2013ApJ...770...13W,2015ApJ...801..123W,2017ApJ...850..108W,2018PASJ...70...36I}. 

Those high resolution studies have provided dynamical masses of the host galaxies. 
They have found that $z \gtrsim 6$ luminous quasars have ratios of SMBH mass to host galaxy mass 
$\sim 10$ times larger than the $z \sim 0$ relation, implying that these SMBHs formed significantly earlier than their hosts. 
However, there may be an observational bias, whereby more luminous quasars are powered 
by more massive SMBHs at high redshifts. 
This affects how closely these observations trace the underlying SMBH mass function at $z \gtrsim 6$, 
if there is a large scatter in $M_{\rm BH}$ for a given galaxy mass \citep[e.g.,][]{2007ApJ...670..249L,2014MNRAS.438.3422S}. 
Indeed, early observations of low-luminosity CFHQS quasars 
showed that they are powered by less massive SMBHs ($\sim 10^8~M_\odot$), 
and show SMBH mass to bulge mass ratios roughly consistent with local galaxies 
\citep{2013ApJ...770...13W,2015ApJ...801..123W,2017ApJ...850..108W}. 
Thus, to achieve a comprehensive view on the early co-evolution of SMBHs and galaxies, 
observations of less-luminous (or smaller $M_{\rm BH}$) quasars are needed. 

With this in mind, we have conducted 
ALMA pilot observations of several optically low-luminosity quasars 
discovered in an on-going deep multi-band ($g, r, i, z, y$), wide area imaging survey \citep{2018PASJ...70S...4A} 
with the Hyper Suprime-Cam \citep[HSC,][]{2018PASJ...70S...1M,Komiyama18,Kawanomoto18,Furusawa18} mounted on the 8.2 m Subaru telescope. 
We have discovered more than 80 low-luminosity quasars 
at $z \gtrsim 6$ down to $M_{\rm 1450} \sim -22$ mag in this survey thus far 
\citep{2016ApJ...828...26M,2018PASJ...70S..35M,2018ApJS..237....5M}, 
including one $z > 7$ object \citep{2019ApJ...872L...2M}. 
Most of these quasars constitute the break of the $z \sim 6$ quasar luminosity function \citep{2018ApJ...869..150M}, 
indicating that they represent the bulk of the quasar population at that high redshift. 
We then organized an intensive multi-wavelength follow-up consortium: 
{\it Subaru High-z Exploration of Low-Luminosity Quasars (SHELLQs)}. 
In \citet{2018PASJ...70...36I}, we presented Cycle 4 ALMA observations 
toward four $z \gtrsim 6$ HSC quasars in the \cii line and 1.2 mm continuum. 
The \cii and continuum luminosities of those HSC quasars are both comparable to local luminous infrared galaxy (LIRG)-class objects, 
suggesting that most of the quasar-host galaxies are less extreme starburst objects than has been found for luminous quasars. 
Like the CFHQS quasars, our HSC quasars tend to show SMBH mass to galaxy mass ratios 
similar to, or even lower than, the local co-evolution relation, which is also a clear contrast to their luminous counterparts. 

In this paper, we report our ALMA Cycle 5 observations of \cii and underlying FIR continuum 
emission towards another three low-luminosity HSC quasars. 
We describe the observations in \S~\ref{sec2}. 
The basic observed properties of both the \cii line and the underlying FIR continuum emission are given in \S~\ref{sec3}. 
We then discuss [C\,\emissiontype{II}]/FIR luminosity ratio, as an ISM diagnostic, 
and the less-biased early co-evolution of SMBHs and galaxies in \S~\ref{sec4}.  
Our findings are basically consistent with our previous Cycle 4 work \citep{2018PASJ...70...36I}: we summarize them in \S~\ref{sec5}. 
Throughout the paper, we adopt the cosmological parameters 
$H_0 = 70$ km s$^{-1}$ Mpc$^{-1}$, $\Omega_{\rm M} = 0.3$, and $\Omega_{\rm \Lambda} = 0.7$.

\section{Observations and data reduction}\label{sec2}
Three $z > 6$ HSC quasars were observed during ALMA Cycle 5 
(ID = 2017.1.00541.S, PI: T. Izumi) at band 6 between 2018 March 20 and 26. 
Our observations are summarized in Table \ref{tbl1}, 
along with the basic target information. 
These observations were each conducted in a single pointing with $\sim 25\arcsec$ diameter field of view (FoV), 
which corresponds to $\sim 140$ kpc at the source redshifts 
(1$\arcsec$ $\sim$ 5.6 kpc). 
The phase tracking centers were set to the optical quasar locations \citep{2018PASJ...70S..35M}. 
The absolute positional uncertainty is $\sim 0''.1$ 
according to the ALMA Knowledgebase\footnote{https://help.almascience.org/index.php?/Knowledgebase/List}. 
With the minimum baseline length (15.1 m), 
the maximum recoverable scales of our observations are $\sim 9.5\arcsec$. 

The receivers were tuned to cover the redshifted \cii line emission, 
whose frequencies were estimated from the measured redshifts of Ly$\alpha$. 
For the J2228$+$0152 observations, the total bandwidth was $\sim 7.5$ GHz, 
divided into four spectral windows of 1.875 GHz width. 
For the J1208$-$0200 and J2239$+$0207 observations, we set three spectral windows 
(i.e., 1.875 $\times$ 3 $\sim$ 5.6 GHz width in total) in one sideband, 
given the large uncertainties of their Ly$\alpha$ redshifts. 
The native spectral resolution was 7.813 MHz (8.7--8.9 km s$^{-1}$), 
but 11--12 channels were binned to improve the signal-to-noise ratio ($S/N$), 
resulting in a final common velocity resolution of $\simeq 100$ km s$^{-1}$. 

Reduction and calibration of the data were performed 
with the Common Astronomy Software Applications package 
\citep[CASA,][]{2007ASPC..376..127M} version 5.1.1, in the standard manner. 
All images were reconstructed with the CASA task \verb|clean| with the Briggs weighting (gain = 0.1, robust = 0.5). 
The achieved synthesized beams and rms sensitivities 
at a velocity resolution of $\sim$100 km s$^{-1}$ are summarized in Table \ref{tbl1}. 
All channels free of line emission were averaged 
to generate a continuum map for each source (see also Table \ref{tbl1}), 
which was subtracted in the $uv$ plane before making the line cube. 
Throughout the paper, only statistical errors are displayed unless otherwise mentioned. 
The systematic uncertainty of the absolute flux calibration at ALMA band 6 is 10\%, 
according to the ALMA Cycle 5 Proposer's Guide. 

\begin{longtable}{*{4}{c}}
\caption{Description of our sample and the ALMA observations}
\label{tbl1}
\hline\hline
 & J1208$-$0200 & J2228$+$0152 & J2239$+$0207 \\
\hline
\endhead
\hline
\endfoot
\hline
\multicolumn{4}{l}{{\bf Note.} Rest-frame UV properties are quoted from \citet{2018PASJ...70S..35M} and \citet{Onoue19}.}\\ 
\multicolumn{4}{l}{The coordinates are updated after tying astrometric calibrations to the {\it Gaia} database.}\\ 
\multicolumn{4}{l}{$^\dag$The Mg\,\emissiontype{II} redshifts are 6.148 (J1208$-$0200) and 6.246 (J2239$+$0207), respectively \citep{Onoue19}.}\\ 
\multicolumn{4}{l}{We did not obtain Mg\,\emissiontype{II} these measurements in hand at the time of our ALMA observations.}\\ 
\endlastfoot
RA (J2000.0) & \timeform{12h08m59s.22} & \timeform{22h28m47s.71} & \timeform{22h39m47s.48} \\ 
Dec (J2000.0) & $-$\timeform{02D00'34''.9} & $+$\timeform{01D52'40''.4} & $+$\timeform{02D07'47''.4} \\
$z_{\rm Ly\alpha}$ & 6.2$^\dag$ & 6.08 & 6.26$^\dag$ \\
$M_{\rm 1450}$ (mag) & $-$24.3 & $-$24.0 & $-$24.6 \\ \hline
Number of antennas & 44--46 & 44--47 & 46--47 \\ 
Baseline (m) & 15.1--783.5 & 15.1--783.5 & 15.1--783.5 \\ 
On-source time (minutes) & 87 & 87 & 65 \\ 
Bandpass calibrator & J1229$+$0203 & J2148$+$0657 & J2148$+$0657 \\
Complex gain calibrator & J1218$-$0119 & J2226+0052 & J2226+0052 \\
Flux calibrator & J1229+0203 & J2148+0657 & J2148+0657 \\ \hline
\multicolumn{4}{c}{\cii cube}\\ \hline
Beam size & 0$\arcsec$.48 $\times$ 0$\arcsec$.39 & 0$\arcsec$.44 $\times$ 0$\arcsec$.40 & 0$\arcsec$.45 $\times$ 0$\arcsec$.38 \\ 
Position Angle (East of North) & $-$59$\arcdeg$.6 & $-$74$\arcdeg$.2 & $-$83$\arcdeg$.2 \\
rms noise per 100 km s$^{-1}$ & \multirow{2}{*}{0.10} & \multirow{2}{*}{0.10} & \multirow{2}{*}{0.11} \\ 
(mJy beam$^{-1}$) &  &  & \\ 
rms noise per 100 km s$^{-1}$ & \multirow{2}{*}{0.15} & \multirow{2}{*}{0.15} & \multirow{2}{*}{0.17} \\ 
(mJy; 1$\arcsec$.0 aperture) &  &  & \\ \hline
\multicolumn{4}{c}{Continuum map}\\ \hline
Observed continuum frequency (GHz) & 266.0 & 260.1 & 261.4 \\
Beam size & 0$\arcsec$.48 $\times$ 0$\arcsec$.38 & 0$\arcsec$.45 $\times$ 0$\arcsec$.41 & 0$\arcsec$.45 $\times$ 0$\arcsec$.38 \\
Position Angle (East of North) & $-$60$\arcdeg$.7 & $-$81$\arcdeg$.1 & $-$82$\arcdeg$.4 \\ 
rms noise & \multirow{2}{*}{16.3} & \multirow{2}{*}{11.2} & \multirow{2}{*}{19.1} \\ 
($\mu$Jy beam$^{-1}$) &  &  & \\
rms noise & \multirow{2}{*}{19.5} & \multirow{2}{*}{15.7} & \multirow{2}{*}{26.0} \\ 
($\mu$Jy; 1$\arcsec$.0 aperture) &  &  & \\

\end{longtable}

\section{Results}\label{sec3}
\begin{figure*}[h]
\begin{center}
\includegraphics[width=\linewidth]{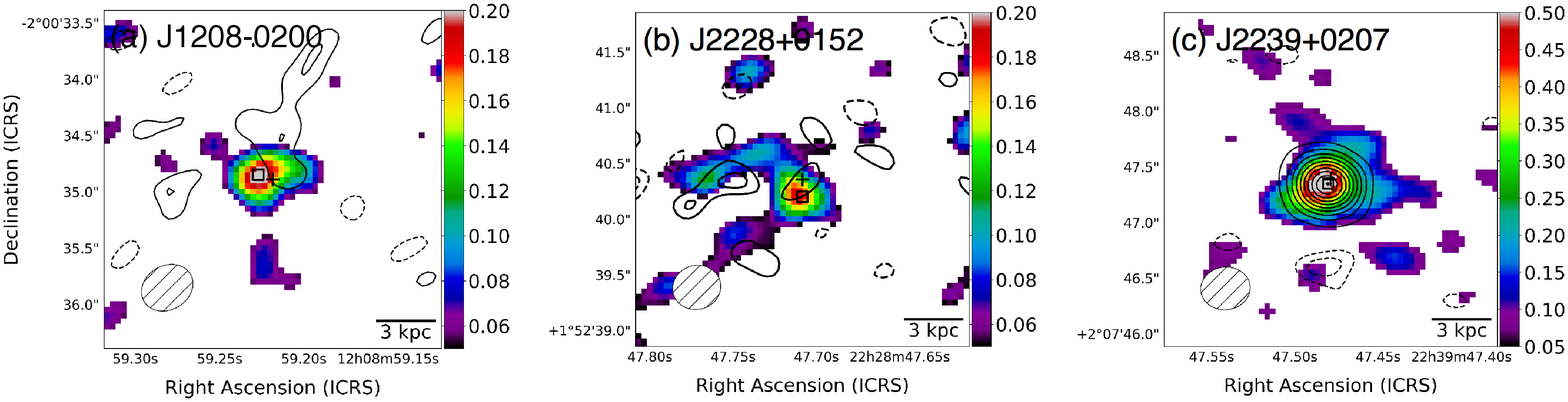}
\end{center}
\caption{
Spatial distributions of the velocity-integrated \cii line (color scale in units of Jy km s$^{-1}$) and rest-frame FIR continuum (contours) emission 
of (a) J1208$-$0200, (b) J2228+0152, and (c) J2239+0207, respectively, shown at the original angular resolutions (see Table \ref{tbl1}). 
The plus and the square symbols denote the optical quasar locations \citep{2018PASJ...70S..35M} and \cii peak locations, respectively. 
Contours indicate: (a) $-$2$\sigma$, 2$\sigma$, 3$\sigma$ (1$\sigma$ = 16.3 $\mu$Jy beam$^{-1}$), 
(b) $-$3$\sigma$, $-$2$\sigma$, 2$\sigma$, 3$\sigma$ (1$\sigma$ = 11.2 $\mu$Jy beam$^{-1}$), 
(c) $-$3$\sigma$, $-$2$\sigma$, 5$\sigma$, 10$\sigma$, $\cdots$, 50$\sigma$ (1$\sigma$ = 19.1 $\mu$Jy beam$^{-1}$). 
Negative values are indicated by the dashed contours. 
The 1$\sigma$ rms sensitivities of the \cii maps are, (a) 0.028, (b) 0.034, and (c) 0.047 Jy beam$^{-1}$ km s$^{-1}$, respectively. 
Pixels below these 1$\sigma$ levels are masked. 
The synthesized beam of the \cii cube is shown in the bottom-left corner of each panel. 
Attenuation due to the primary beam patterns are not corrected. 
\label{fig1}} 
\end{figure*}

Figure \ref{fig1} displays the spatial distribution of the velocity-integrated \cii line emission (0th moment) 
as well as the underlying rest-frame FIR continuum emission ($\lambda_{\rm rest}$ $\simeq$ 158 $\mu$m) 
of the three objects observed in Cycle 5. 
Those moment 0 maps were made with the CASA task \verb|immoments|, 
integrating over the full velocity range containing the line emission
\footnote{The \cii emission line of J2239+0207 spans over two spectral windows (see Figure \ref{fig2}). 
We thus integrated the emission between 261.50 -- 262.25 GHz in one window and 262.25 -- 262.75 GHz in the other, and combined them to generate the moment 0 map.}. 
The \cii emission was detected in all sources, with no apparent spatial offset of their peak locations 
from the optical centroids given the astrometric uncertainties. 
However, FIR continuum emission was only significantly detected in J2239+0207 at the original angular resolution. 
Given that the \cii emission seems to be somewhat extended relative to the synthesized beams,  
we measured FIR properties of these quasar host galaxies within a common 1\arcsec.0 aperture. 
With this treatment, FIR continuum emission was marginally detected in J1208$-$0200 as well. 
The rms sensitivities for these 1\arcsec.0 aperture measurements are listed in Table \ref{tbl1}. 
The relevant FIR properties of the targets are shown in Table \ref{tbl2}, 
which will be explained in detail in the following.

\subsection{\cii line properties}\label{sec3.1}
\begin{figure}[h]
\begin{center}
\includegraphics[width=\linewidth]{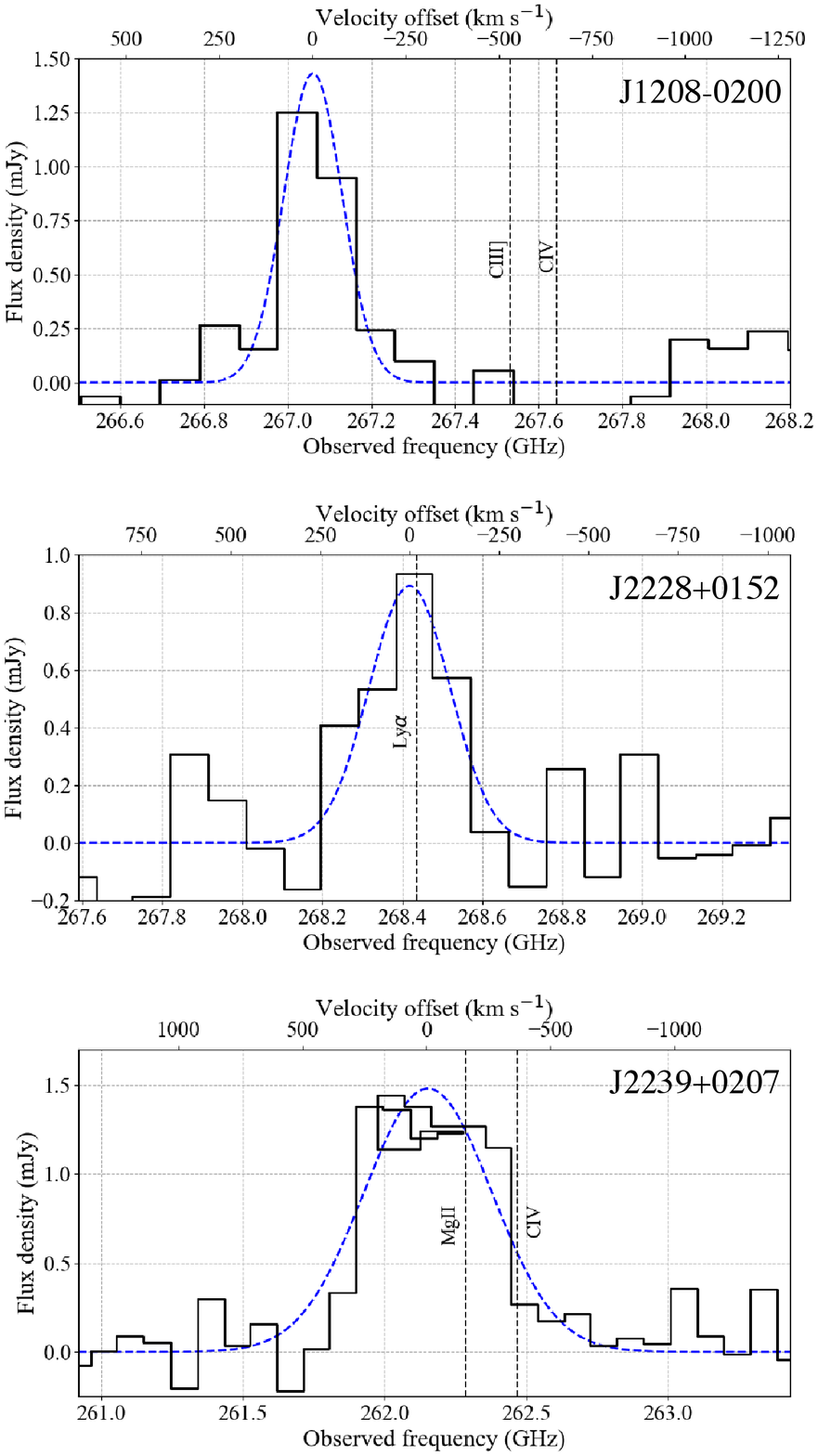}
\end{center}
\caption{
\cii spectra (black solid line) of the three HSC quasars observed with ALMA, 
along with our best-fit single Gaussian profiles (blue dashed line). 
These were measured with 1$\arcsec$.0 aperture placed 
either at the peak location of the FIR continuum emission (J2239+0207) 
or the rest-UV quasar position (J1208$-$0200 and J2228+0152; 
FIR continuum emission was not detected at the original resolutions in these objects). 
The redshifts determined from the rest-UV emission lines \citep{Onoue19} are indicated by the vertical dashed lines. 
Note that the measured Mg\,\emissiontype{II} line of J1208$-$0200 
is significantly redshifted with respect to this \cii line, which is out of the displayed range. 
The \cii emission of J2239+0207 spans over two spectral windows. 
Thus there are overlapped channels at a velocity offset of $\sim 0$ km s$^{-1}$. 
\label{fig2}} 
\end{figure}

Figure \ref{fig2} displays the \cii line spectra measured with the 1\arcsec.0 aperture. 
Given the modest signal-to-noise ratio (S/N) achieved, we simply fitted each spectrum with a single Gaussian profile, 
which delivered the velocity centroid (or redshift = $z_{\rm \cii}$), line width (full width at half maximum = FWHM$_{\rm \cii}$), 
and the velocity-integrated line flux ($S_{\rm \cii}$) of the quasar host galaxy (Table \ref{tbl2}). 

\begin{figure}
\begin{center}
\includegraphics[width=\linewidth]{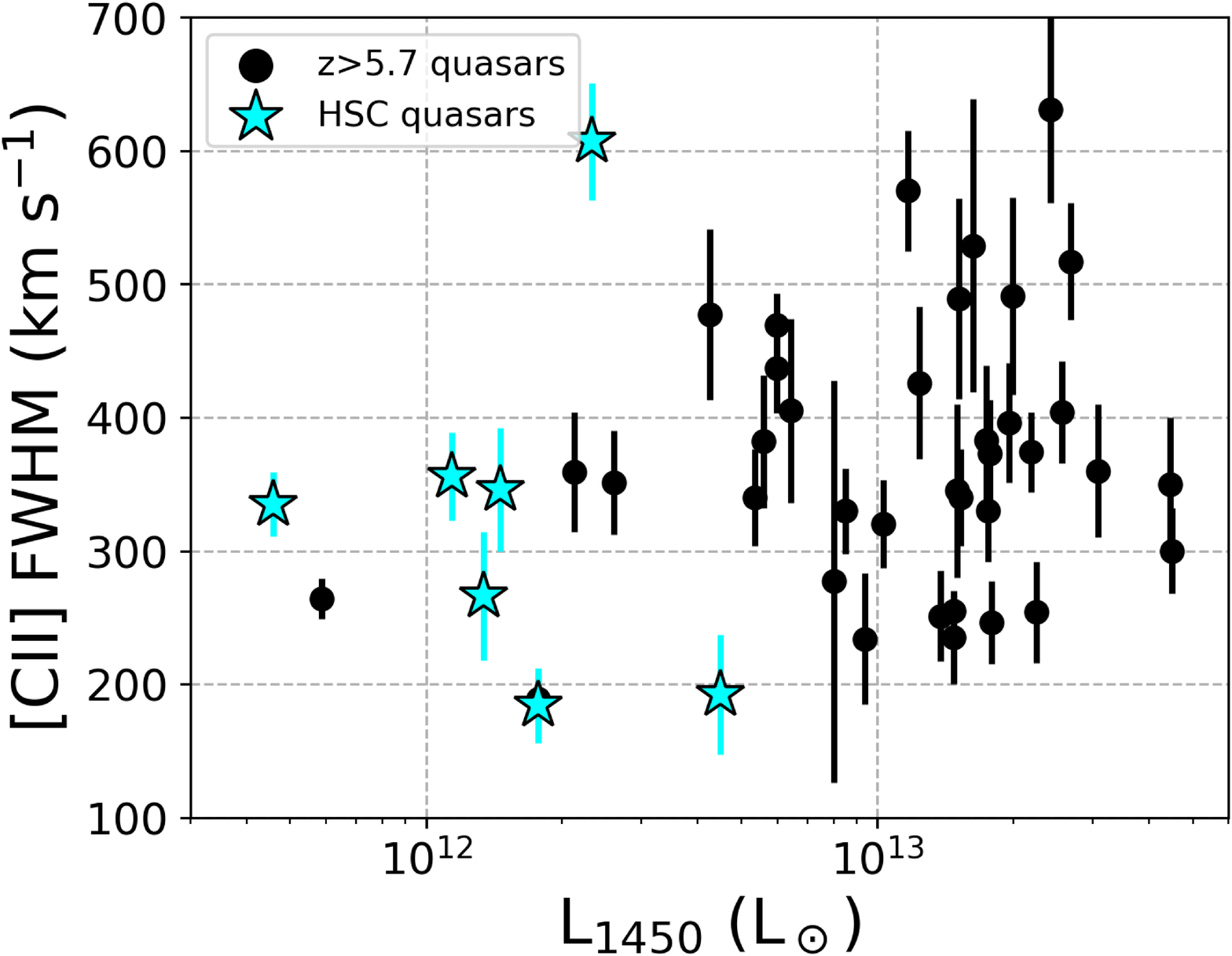}
\end{center}
\caption{
Gaussian FWHM of \cii line as a function of the quasar absolute UV magnitude ($M_{\rm 1450}$). 
Compilations of $z > 5.7$ quasars from the literature 
\citep[circle,][]{2005A&A...440L..51M,2018ApJ...854...97D,
2012ApJ...751L..25V,2016ApJ...816...37V,2017ApJ...851L...8V,
2013ApJ...770...13W,2015ApJ...801..123W,2017ApJ...850..108W,
2015ApJ...805L...8B,2017ApJ...849...91M} 
are plotted along with the HSC quasars (star) measured in \citet{2018PASJ...70...36I} and in this work. 
\label{fig_add1}} 
\end{figure}

The FWHM$_{\rm \cii}$ of these three HSC quasar host galaxies presented here, 
as well as those measured in our Cycle 4 work \citep{2018PASJ...70...36I}
\footnote{The mean and standard deviation of the distribution is 327 $\pm$ 135 km s$^{-1}$ for the full sample of seven objects (Cycle 4 + 5).}, 
are consistent with the distribution constructed from a large sample of $z \gtrsim 6$ quasar host galaxies \citep{2018ApJ...854...97D}, 
but J1208$-$0200 and J2239+0207 lie at the lower and higher extremes of the distribution, respectively (Figure \ref{fig_add1}). 
Thus, there seems to be no clear correlation of FWHM$_{\rm \cii}$ and quasar luminosity: 
indeed, the Spearman rank correlation coefficient for the relation in Figure \ref{fig_add1} 
is only $-$0.21 with a null hypothesis probability of 0.16. 
The line profile of J2239+0207 is clearly flat at the peak, 
in the velocity range of $-$300 to +300 km s$^{-1}$, 
although we fit the profile with a single Gaussian for simplicity. 
A flat line profile was also found in a $z = 4.6$ quasar \citep{2015MNRAS.452...88K} 
and a $z = 6.13$ quasar \citep{2017ApJ...845..138S}. 
Such a profile suggests that the \cii line emission originates from a rotating disk. 
On the other hand, we cannot discuss the detailed 
dynamical nature (e.g., rotation-dominant or dispersion-dominant) 
of J1208$-$0200 and J2228$+$0152 as these are barely resolved. 

The velocity centroid ($z_{\rm \cii}$) of J2228+0152 agrees well with $z_{\rm Ly\alpha}$ (Table \ref{tbl1}), 
but $z_{\rm \cii}$ is offset significantly blueward from $z_{\rm Ly\alpha}$ for J1208$-$0200 and J2239+0207. 
These offsets could simply be the consequence of the considerable uncertainties 
in $z_{\rm Ly\alpha}$ due to severe intergalactic absorption \citep[$\Delta z \sim 1000$ km s$^{-1}$, e.g.,][]{2017ApJ...840...24E}. 
Among the ionized lines predominantly emerging from the broad line region of quasars, 
C\,\emissiontype{IV} $\lambda$1549 is usually found to be blueshifted 
with respect to the host galaxies \citep{2016ApJ...831....7S,2017ApJ...849...91M}. 
In the case of our 7 HSC quasars with \cii measurements, the measured blueshifts of C\,\emissiontype{IV} 
are $\sim 400 - 600$ km s$^{-1}$ with respect to $z_{\rm \cii}$ \citep{Onoue19}. 
The blueshifted nature of C\,\emissiontype{IV} indicates the existence of outflowing gas close to the central SMBHs. 

\begin{figure}
\begin{center}
\includegraphics[width=\linewidth]{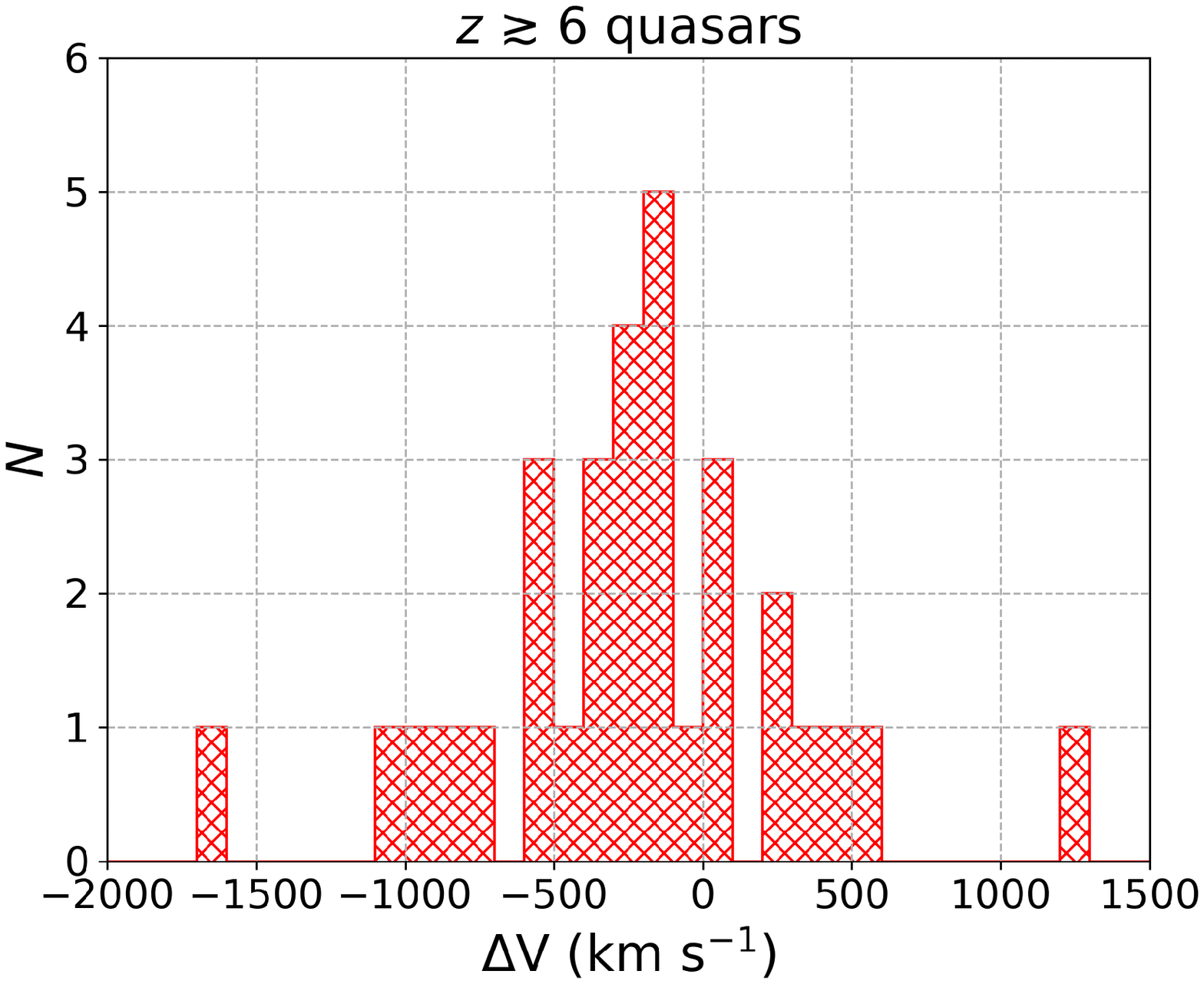}
\end{center}
\caption{
Histogram of the velocity offset 
between Mg\,\emissiontype{II}-based redshift 
and [C\,\emissiontype{II}]-based redshift ($\Delta v = v_{\rm Mg\,\emissiontype{II}} - v_{\rm \cii}$) 
for $z > 5.7$ quasars compiled from the literature 
\citep[][]{2005A&A...440L..51M,2018ApJ...854...97D,
2012ApJ...751L..25V,2016ApJ...816...37V,2017ApJ...851L...8V,2013ApJ...773...44W,2016ApJ...830...53W,
2013ApJ...770...13W,2015ApJ...801..123W,2017ApJ...850..108W,2015ApJ...805L...8B,2017ApJ...849...91M}, 
along with our HSC quasars. 
The Mg\,\emissiontype{II} line is predominantly blueshifted in these $z \gtrsim 6$ quasars 
with a mean and standard deviation of $-$284 $\pm$ 607 km s$^{-1}$. 
\label{fig_rev1}} 
\end{figure}

Regarding Mg\,\emissiontype{II} $\lambda$2798, however, 
we do not see significant blueshifts ($\gtrsim 500$ km s$^{-1}$) in these HSC quasars \citep{Onoue19}
\footnote{J1208$-$0200 even shows a large redshift in Mg\,\emissiontype{II} with respect to 
\cii by $\sim 1260$ km s$^{-1}$ (out of the displayed range in Figure \ref{fig2}), 
although its Mg\,\emissiontype{II} profile would be affected by OH sky emission \citep{Onoue19}.}, 
while some other quasars show such blueshifts 
\citep[e.g.,][]{2016ApJ...816...37V,2016ApJ...830...53W,2018ApJ...854...97D}, 
as shown in the compilation of the velocity shift ($\Delta v = v_{\rm Mg\,\emissiontype{II}} - v_{\rm \cii}$, Figure \ref{fig_rev1}). 
The mean and standard deviation of the distribution is $-$284 $\pm$ 607 km s$^{-1}$. 
Although nuclear outflows can be a possible origin of such shifts, 
only a marginal and no clear correlations are found for $\Delta v$, 
between quasar nuclear luminosity or Eddington ratio
\footnote{We used the bolometric correction factor of 4.4 from 1450 {\AA} luminosity \citep{2006ApJS..166..470R} 
and Mg\,\emissiontype{II}-based $M_{\rm BH}$ data to compute the Eddington ratios.}, 
with the corresponding Spearman rank coefficient of 0.35 ($p$-value = 0.05) and 0.04 ($p$-value = 0.82), 
respectively, for the samples shown in Figure \ref{fig_rev1}. 
Hence the cause of the Mg\,\emissiontype{II} blueshifts remains unclear. 
It is nevertheless advisable to use the Mg\,\emissiontype{II} line with caution 
as a good indicator of the systemic redshift, given the wide distribution of $\Delta v$. 
This is particularly an issue for ALMA observations, 
as the width of its single baseband is only $\sim 2250$ km s$^{-1}$ at 250 GHz.

\begin{figure}
\begin{center}
\includegraphics[width=\linewidth]{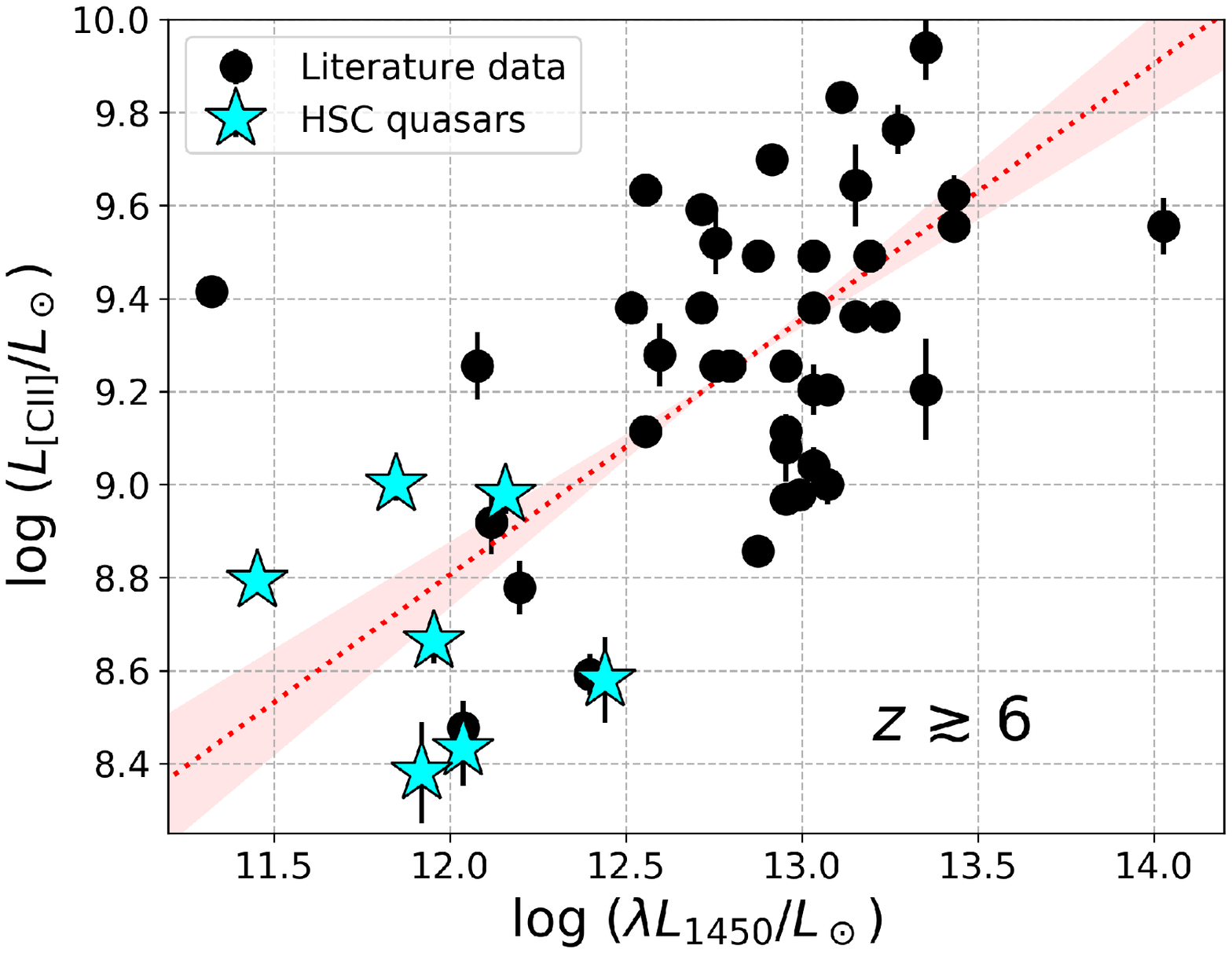}
\end{center}
\caption{
Relationship of the quasar UV luminosity ($\lambda L_{\rm 1450}$) 
and \cii line luminosity ($L_{\rm \cii}$) shown in the logarithmic scale. 
Literature data of $z \gtrsim 6$ quasars are compiled 
\citep[circle,][]{2005A&A...440L..51M,2018ApJ...854...97D,
2012ApJ...751L..25V,2016ApJ...816...37V,2017ApJ...851L...8V,2013ApJ...773...44W,2016ApJ...830...53W,
2013ApJ...770...13W,2015ApJ...801..123W,2017ApJ...850..108W,2015ApJ...805L...8B,2017ApJ...849...91M}. 
Also plotted are the HSC quasars (star). 
The dotted line (red) and the shaded region indicate our best fit 
linear regression line and its 1$\sigma$ scatter, respectively.}
\label{fig_rev2}
\end{figure}

The \cii line luminosities of our sources were calculated with the standard equation, 
$L_{\rm \cii} = 1.04 \times 10^{-3}~S_{\rm \cii}~\nu_{\rm rest}~(1+z_{\rm \cii})^{-1}~D^2_L$ \citep{2005ARA&A..43..677S}, 
where $L_{\rm \cii}$ is the \cii line luminosity in units of $L_\odot$ and $D_L$ is the luminosity distance in units of Mpc. 
The resultant luminosities lie in the range $L_{\rm \cii} = (2.4-9.5) \times 10^8~L_\odot$, 
consistent with our Cycle 4 measurements for the other four HSC quasars 
\citep[$L_{\rm \cii} \simeq (4-10) \times 10^8~L_\odot$,][]{2018PASJ...70...36I}. 
These values are also comparable to other low-luminosity quasars 
at $z \gtrsim 6$ \citep{2013ApJ...770...13W,2015ApJ...801..123W,2017ApJ...850..108W}. 
In contrast, optically-luminous quasars at $z \gtrsim 6$ show higher $L_{\rm \cii}$, 
typically in the range $\simeq (1-10) \times 10^9~L_\odot$ \citep[e.g.,][]{2013ApJ...773...44W,2016ApJ...830...53W,2016ApJ...816...37V,2018ApJ...854...97D}. 
Indeed there is a positive correlation between 
the quasar nuclear luminosity (represented by the 1450 {\AA} monochromatic luminosity) 
and $L_{\rm \cii}$ as shown in Figure \ref{fig_rev2}. 
The derived Spearman rank coefficient is 0.59, with the associated $p$-value of 2.0 $\times$ 10$^{-5}$. 
Our linear regression analysis returns the best fit line for the correlation as 
\begin{equation}
\log \left( \frac{L_{\rm \cii}}{L_\odot} \right) = (2.21 \pm 1.13) + (0.55 \pm 0.09) \times \log \left( \frac{\lambda L_{\rm 1450}}{L_\odot} \right).
\end{equation} 
As the FIR emission is likely to be dominated by cold dust emission 
powered by star formation \citep[e.g.,][]{2014ApJ...785..154L}, 
one plausible and simple explanation for the above correlation is that 
both the black hole accretion and the star formation is driven by a common reservoir of gas. 

The SFR of these quasars can then be estimated by attributing the \cii emission heating solely 
to young stars: ${\rm SFR}_{\rm \cii}/M_\odot~{\rm yr^{-1}} = 1.0 \times 10^{-7} (L_{\rm \cii}/L_\odot)^{0.98}$ \citep{2011MNRAS.416.2712D}. 
This relation has an intrinsic scatter of 0.3 dex and is based on the Kroupa initial mass function \citep[IMF,][]{2001MNRAS.322..231K}. 
The derived values range from 16 to 63 $M_\odot$ yr$^{-1}$ (Table \ref{tbl2}), 
well within the SFR-range of local LIRG-class systems \citep[e.g.,][]{2013ApJ...774...68D,2014ApJ...790...15S}. 
Note that the \citet{2011MNRAS.416.2712D} relation was derived for objects with $L_{\rm FIR} \lesssim 10^{12}~L_\odot$, 
thus it may not be appropriate to apply this for J2239+0207 ($L_{\rm FIR} \simeq 2 \times 10^{12}~L_\odot$). 

\begin{longtable}{*{4}{c}}
\caption{Rest-frame FIR properties of the HSC quasars \label{tbl2}}
\hline
\hline
 & J1208$-$0200 & J2228+0152 & J2239+0207 \\
\hline
\endhead
\hline 
\endfoot
\hline
\multicolumn{4}{l}{{\bf Note.} These were measured with a common 1$\arcsec$.0 aperture.}\\
\multicolumn{4}{l}{The FIR luminosities were estimated with a graybody spectrum model.}\\
\multicolumn{4}{l}{The upper limits are the 3$\sigma$ values.}\\
\multicolumn{4}{l}{${\rm SFR}_{\rm \cii}$/$M_\odot$~yr$^{-1}$ = 1.0 $\times$ 10$^{-7}$ ($L_{\rm [C\,\emissiontype{II}]}$/$L_\odot$)$^{0.98}$ \citep{2011MNRAS.416.2712D}.}\\
\multicolumn{4}{l}{${\rm SFR}_{\rm TIR}/M_\odot~{\rm yr^{-1}} = 1.49 \times 10^{-10} L_{\rm TIR}/L_\odot$ \citep{2011ApJ...737...67M}.}\\
\endlastfoot
$z_{\rm \cii}$ & 6.1165 $\pm$ 0.0002 & 6.0805 $\pm$ 0.0004 & 6.2497 $\pm$ 0.0004 \\
FWHM$_{\rm \cii}$ (km s$^{-1}$) & 184 $\pm$ 28 & 266 $\pm$ 48 & 607 $\pm$ 44 \\ 
$S_{\rm \cii}$ (Jy km s$^{-1}$) & 0.280 $\pm$ 0.056 & 0.253 $\pm$ 0.059 & 0.955 $\pm$ 0.085 \\ 
$L_{\rm{\cii}}$ (10$^8$ $L_\odot$) & 2.71 $\pm$ 0.54 & 2.43 $\pm$ 0.57 & 9.53 $\pm$ 0.85 \\ 
$f_{\rm 1.2mm}$ ($\mu$Jy) & 85 $\pm$ 20 & $<$47 & 1110 $\pm$ 26 \\ 
EW$_{\rm{\cii}}$ ($\mu$m) & 1.73 $\pm$ 0.52 & $>$2.82 & 0.45 $\pm$ 0.04 \\
SFR$_{\rm \cii}$ ($M_\odot$ yr$^{-1}$) & 18 $\pm$ 4 & 16 $\pm$ 5 & 63 $\pm$ 8 \\ \hline
\multicolumn{4}{c}{Assumption: $T_{\rm d}$ = 47 K, $\beta$ = 1.6, $\kappa_{\rm 250}$ = 0.4 cm$^2$ g$^{-1}$}\\ \hline
$L_{\rm FIR}$ (10$^{11}$ $L_\odot$) & 1.62 $\pm$ 0.37 & $<$0.94 & 21.74 $\pm$ 0.51 \\
$L_{\rm TIR}$ (10$^{11}$ $L_\odot$) & 2.29 $\pm$ 0.52 & $<$1.34 & 30.65 $\pm$ 0.72 \\
SFR$_{\rm TIR}$ ($M_\odot$ yr$^{-1}$) & 34 $\pm$ 8  & $<$20 & 453 $\pm$ 10 \\
$M_{\rm dust}$ ($10^7~M_\odot$) & 1.2 $\pm$ 0.3 & $<$0.7 & 15 $\pm$ 1 \\
$L_{\rm \cii}/L_{\rm FIR}$ ($10^{-3}$) & 1.67 $\pm$ 0.51 & $>$2.58 & 0.44 $\pm$ 0.04 \\
\end{longtable}

We used the CASA task \verb|imfit| to fit a two-dimensional Gaussian profile 
to the \cii integrated intensity (0$^{\rm th}$ moment) maps, 
and estimated their beam-deconvolved spatial extents. 
The maps made with the original resolutions (Figure \ref{fig1}) were used for this purpose. 
This image-plane fitting method has been widely used 
in previous submm studies of $z \gtrsim 6$ quasar host galaxies \citep[e.g.,][]{2015ApJ...801..123W,2016ApJ...816...37V}, 
which enables a direct comparison with these earlier studies. 
The estimated values are listed in Table \ref{tbl3}: their FWHM sizes are $\sim 2.1-4.0$ kpc (major axis). 
Although the associated uncertainties are admittedly large, 
these sizes are comparable to those found in our previous work 
on the other four HSC quasars, as well as to many optically luminous quasars \citep{2018PASJ...70...36I}.

\begin{longtable}{*{3}{c}}
\caption{Spatial extent of the star-forming region}\label{tbl3}
\hline
\hline
Name & Size ([C\,\emissiontype{II}] FWHM) & Size (FIR continuum FWHM) \\ \hline
\endhead
\hline 
\endfoot
\hline
\multicolumn{3}{l}{{\bf Note.} The original resolution data was used for the measurements.}\\
\endlastfoot
\multirow{2}{*}{J1208$-$0200} & (0\arcsec.63 $\pm$ 0\arcsec.06) $\times$ (0\arcsec.35 $\pm$ 0\arcsec.07) & \multirow{2}{*}{-} \\
 & (3.6 $\pm$ 0.3) kpc $\times$ (1.9 $\pm$ 0.4) kpc & \\ \hline
 \multirow{2}{*}{J2228+0152} & (0\arcsec.38 $\pm$ 0\arcsec.05) $\times$ (0\arcsec.18 $\pm$ 0\arcsec.07) & \multirow{2}{*}{-} \\
 & (2.1 $\pm$ 0.3) kpc $\times$ (1.0 $\pm$ 0.4) kpc & \\ \hline
 \multirow{2}{*}{J2239+0207} & (0\arcsec.72 $\pm$ 0\arcsec.02) $\times$ (0\arcsec.46 $\pm$ 0\arcsec.02) & (0\arcsec.22 $\pm$ 0\arcsec.03) $\times$ (0\arcsec.11 $\pm$ 0\arcsec.05) \\
 & (4.0 $\pm$ 0.1) kpc $\times$ (2.6 $\pm$ 0.1) kpc & (1.2 $\pm$ 0.2) kpc $\times$ (0.7 $\pm$ 0.5) kpc \\ 
\end{longtable}

\subsection{FIR continuum properties}\label{sec3.2}
The observed 1.2 mm continuum flux densities ($f_{\rm 1.2mm}$) are used to determine their FIR luminosities ($L_{\rm FIR}$) 
integrated over the rest-frame wavelengths of $\lambda_{\rm rest} = 42.5 - 122.5$ $\mu$m \citep{1988ApJS...68..151H}. 
Here we assume a graybody spectrum with dust temperature of $T_{\rm d} = 47$ K 
and emissivity index\footnote{Emissivity $\propto \nu^\beta$} of $\beta = 1.6$ 
based on the mean spectral energy distribution of high redshift optically and FIR luminous quasars 
\citep{2006ApJ...642..694B,2014ApJ...785..154L}, 
to be consistent with previous work \citep[e.g.,][]{2013ApJ...773...44W,2013ApJ...770...13W,2016ApJ...816...37V}. 
However, these values are likely to vary significantly from source to source \citep{2018ApJ...866..159V,2019arXiv190210727L}. 
If our HSC quasars instead have $T_{\rm d}$ close to the value found for nearby LIRG-class systems \citep[$\sim 35$ K,][]{2012ApJS..203....9U}, 
the resultant inferred $L_{\rm FIR}$ values would be $\sim 3 \times$ lower. 
We hereafter only consider the uncertainties of flux measurements, not that of the $T_{\rm d}$, 
which should be constrained further with future multi-wavelength observations. 
Note that the influence of the cosmic microwave background (CMB) radiation 
on the submm observations at high redshifts \citep{2013ApJ...766...13D} is not considered, 
as that effect is negligible as long as we adopt $T_{\rm d} = 47$ K.

The $f_{\rm 1.2mm}$ and $L_{\rm FIR}$ measured with the common 1\arcsec.0 aperture are listed in Table \ref{tbl2}. 
J1208$-$0200 was marginally detected ($\sim 4\sigma$) with $L_{\rm FIR} = (1.6 \pm 0.4) \times 10^{11}~L_\odot$, 
which is slightly smaller than those of the four Cycle 4 HSC quasars 
\citep[$L_{\rm FIR} \sim (3-5) \times 10^{11}~L_\odot$,][]{2018PASJ...70...36I}. 
J2228+0152 is undetected, with a $3\sigma$ upper limit of 
$f_{\rm 1.2mm} < 47$ $\mu$Jy and $L_{\rm FIR} < 9 \times 10^{10}~L_\odot$ 
(i.e., below the luminosity range of a LIRG), making it one of the lowest 
$L_{\rm FIR}$ quasar host galaxies ever studied at $z \gtrsim 6$. 
The $L_{\rm FIR}$ of these HSC quasars are thus smaller by factors of $\sim 10-100$ 
than most of the $z \gtrsim 6$ optically-luminous quasars studied thus far \citep[e.g.,][]{2007AJ....134..617W,2008ApJ...687..848W}. 
On the other hand, for J2239+0207 we found $L_{\rm FIR} \simeq 2 \times 10^{12}~L_\odot$, 
showing that there is a broad distribution in $L_{\rm FIR}$ even among HSC quasars 
of comparable UV/optical luminosities. 
The relationship between quasar luminosity and $L_{\rm FIR}$ is further discussed in \S~\ref{sec4.2.2}. 

We measured the size of the FIR continuum-emitting region of J2239+0207 with the \verb|imfit| task, 
finding a significantly smaller size than that of the \cii emitting region (Table \ref{tbl3}; see also Figure \ref{fig1}). 
The \cii emitting region is often larger than the continuum-emitting region in high-$z$ quasars 
\citep[e.g.,][]{2013ApJ...773...44W,2016ApJ...816...37V}, although the cause for this remains unclear. 

The total infrared luminosity ($L_{\rm TIR}$) integrated over the 8--1000 $\mu$m range 
is supposed to be powered by star formation, and thus gives an independent estimate of SFR. 
We use the conversion, ${\rm SFR_{\rm TIR}}/M_\odot~{\rm yr}^{-1} = 1.49 \times 10^{-10}~L_{\rm TIR}/L_\odot$ 
\citep{2011ApJ...737...67M}, which is based on the Kroupa IMF \citep{2001MNRAS.322..231K}. 
The SFRs based on this method are also listed in Table \ref{tbl2}. 
We also derived dust mass $M_{\rm dust}$ from $L_{\rm FIR}$ 
with $M_{\rm dust} = L_{\rm FIR}/(4\pi \int \kappa_\nu B_\nu d\nu)$, 
where $\kappa_\nu$ is the mass absorption coefficient, taken to be $\kappa_\nu = \kappa_0 (\nu/250~{\rm GHz})^\beta$ 
with $\kappa_0 = 0.4$ cm$^2$ g$^{-1}$ \citep{2004A&A...425..109A}, and $B_\nu$ is the Planck function. 
The derived values span a wide range from $<7 \times 10^6~M_\odot$ to $1.5 \times 10^8~M_\odot$ (Table \ref{tbl2}).

\subsection{Further details of J2228+0152 and J2239+0207: interactions and companions?}\label{sec3.3}
\begin{figure}[h]
\begin{center}
\includegraphics[width=\linewidth]{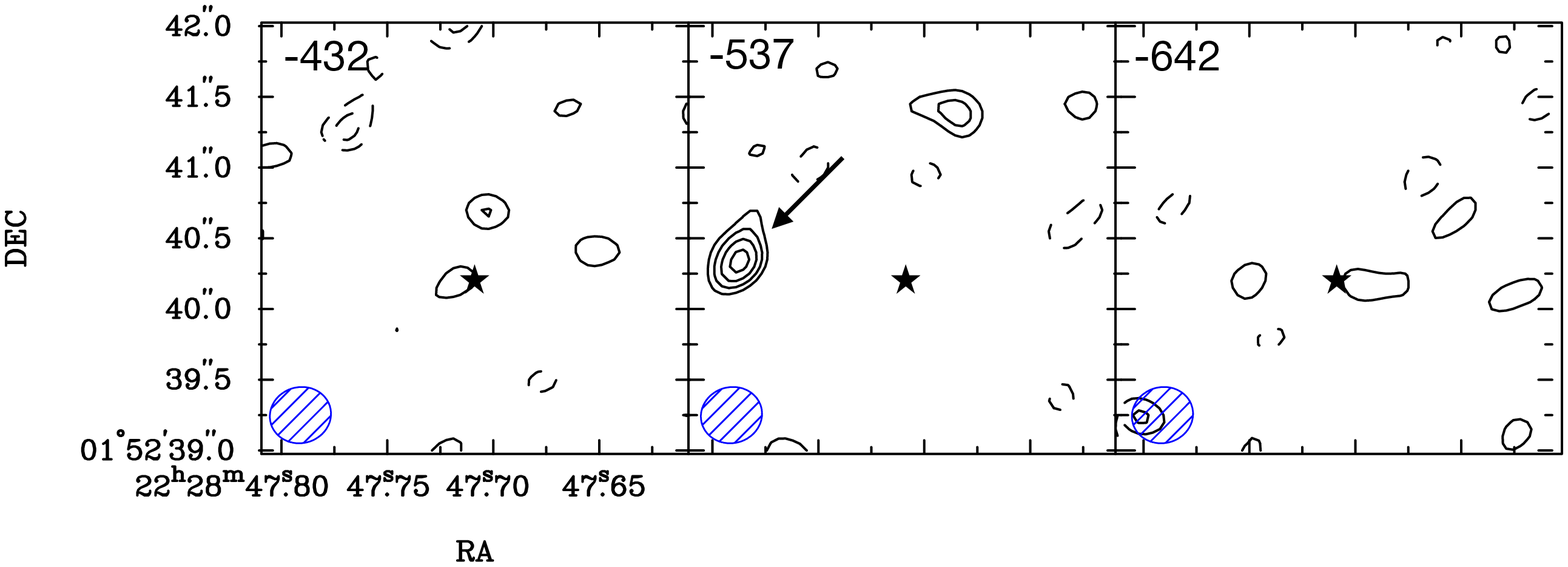}
\end{center}
\caption{
A companion \cii emitter candidate found in the J2228+0152 field. 
These velocity channel maps were generated from the original resolution 
\cii cube (angular resolution = 0\arcsec.44 $\times$ 0\arcsec.40; bottom-left ellipses). 
The candidate emitter is indicated by the arrow, which is $\sim 7$ kpc away from the central quasar (star). 
Contours step as $-$3, $-$2, 2, 3, 4, 5$\sigma$ (1$\sigma$ = 0.10 mJy beam$^{-1}$ at a velocity resolution of $\simeq 100$ km s$^{-1}$). 
The numbers in the top-left corner indicate the relative velocities to the quasar systemic velocity. 
\label{fig_add2}} 
\end{figure}

The \cii emission of J2228+0152 seems to be extended to the east, 
although the statistical significance of the extended component is only modest ($\sim 3-3.5\sigma$). 
However, there is also weak ($\sim 2-3\sigma$) FIR continuum emission around the eastern extension, 
which motivates us to further investigate its structure. 
To this end, we constructed \cii velocity channel maps of J2228+0152 using the MIRIAD software \citep{1995ASPC...77..433S}. 
We found that there is one \cii emitter candidate at the location of 
the eastern extension (Figure \ref{fig_add2}), which is detected at $5.5\sigma$. 
If this is a \cii emitter, it is located at $z_{\rm \cii} = 6.068$ with $S_{\rm \cii} = 0.055$ Jy km s$^{-1}$ or $L_{\rm \cii} = 5.3 \times 10^7~L_\odot$. 
The velocity offset and the projected separation on the sky of this emitter candidate, 
measured from the central quasar, are $\sim -540$ km s$^{-1}$ and $\sim 7$ kpc, respectively. 
Thus, the eastern extension may be related to the interaction of J2228+0152 and this companion emitter. 
Although we need higher sensitivity observations to further study 
the nature of this emitter as it was only detected in one channel, 
companion galaxies and extended (or interacting) morphologies 
have been identified in \cii emission around some quasars at $z \gtrsim 6$ \citep{2017Natur.545..457D}. 

\begin{figure}[h]
\begin{center}
\includegraphics[width=\linewidth]{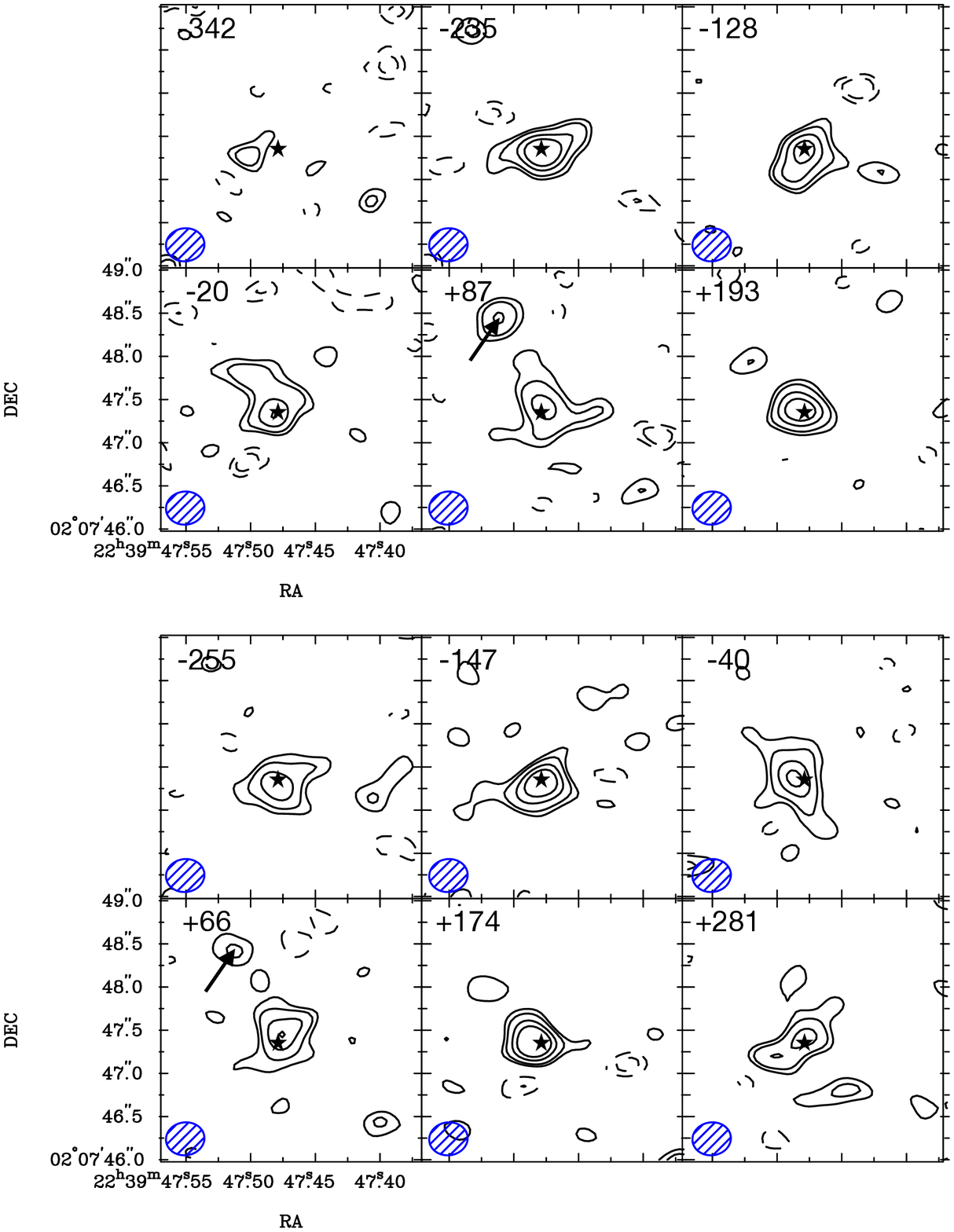}
\end{center}
\caption{
Velocity channel maps of J2239+0207 generated from the original resolution \cii cube 
(i.e., angular resolution = 0\arcsec.45 $\times$ 0\arcsec.38; bottom-left ellipses). 
Contours step as $-$3, $-$2, 2, 3, 5, 7$\sigma$ (1$\sigma$ = 0.11 mJy beam$^{-1}$ at a velocity resolution of $\simeq 100$ km s$^{-1}$). 
The top and bottom sets of panels show the data from different spectral windows, which partially overlap. 
The FIR continuum peak position is marked by the star symbol. 
The location of a candidate companion \cii emitter is indicated by the arrows. 
The number in the top-left corner of each panel indicates the relative velocity to the quasar systemic velocity. 
\label{fig3}} 
\end{figure}

Another interesting object is J2239+0207, as it shows the highest $L_{\rm FIR}$ 
of the seven HSC quasars thus far studied with ALMA. 
One possible origin of the high FIR luminosity is a merger of two or more galaxies, 
as is often observed in nearby ULIRGs \citep[e.g.,][]{1996ARA&A..34..749S}. 
Thus, we searched for features indicative of galaxy mergers or interactions 
in \cii velocity channel maps of J2239+0207 (Figure \ref{fig3}). 
As we remarked in \S~\ref{sec2}, the \cii line spans two spectral windows, 
allowing us to test the robustness of the features we see. 
Note, however, that the velocity spacings of these two data sets are not perfectly matched. 

The \cii emission of J2239+0207 is slightly extended relative to the synthesized beam 
and shows filamentary structures in some channels. 
Although these components are of low statistical significance ($\lesssim 3-5\sigma$) 
and do not seem to be well matched in the two windows, 
we found a \cii emission candidate at exactly the same location in the two datasets, 
at similar velocities (+87 km s$^{-1}$ in one spectral window and +66 km s$^{-1}$ in the other). 
If this is indeed a \cii emitter, an interaction with J2239+0207 (projected separation $\sim 1\arcsec$ $\sim$ 5.6 kpc) 
may have triggered the observed starburst activity. 
Meanwhile, if it is real, this \cii emitter ($z_{\rm \cii} = 6.248$) has $S_{\rm \cii} \sim 0.063$ Jy km s$^{-1}$ 
or $L_{\rm \cii} \sim 6.2 \times 10^7~L_\odot$ (corresponding to ${\rm SFR}_{\rm \cii} \sim 4~M_\odot$ yr$^{-1}$), which are rather modest values. 
Given the compact size and narrow velocity width, a dynamical mass of this emitter should be small. 
Thus a possible future merger of this emitter and J2239+0207 would be rather minor. 
J2239+0207 itself must thus have copious amount of cold material to support the observe starburst activity. 
However, our limited sensitivity (the significance of the emission is $\sim 5\sigma$ and $\sim 3\sigma$ in each window, respectively; 
FIR continuum emission is below 3$\sigma$ at this position) 
means that we cannot call this a robust detection. 
Again, we need much deeper observations to reveal its nature.

\subsection{Continuum emitters}\label{sec3.4} 
\begin{figure}[h]
\begin{center}
\includegraphics[width=\linewidth]{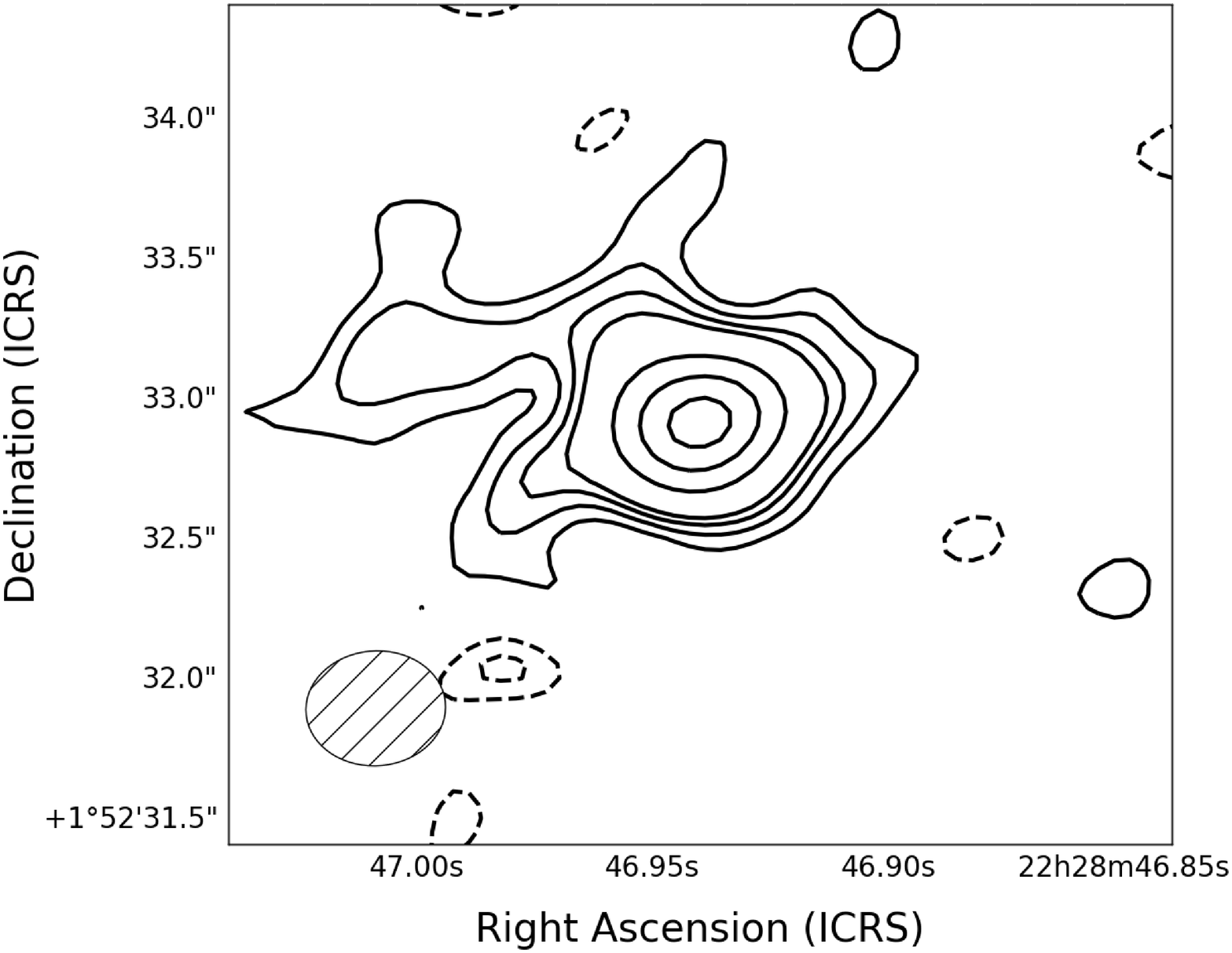}
\end{center}
\caption{
A continuum emitter found in the J2228+0152 field (13\arcsec.7 away from the phase reference position). 
The contours are $-$3$\sigma$, $-$2$\sigma$, 2$\sigma$, 3$\sigma$, 4$\sigma$, 5$\sigma$, 
10$\sigma$, 15$\sigma$, and 20$\sigma$, with 1$\sigma$ = 11.2 $\mu$Jy beam$^{-1}$. 
The emission is spatially resolved with the peak flux is 0.25 mJy beam$^{-1}$, 
and the area-integrated flux is 0.38 $\pm$ 0.05 mJy, respectively. 
The photometric redshift predicted for this object is $z_{\rm photo} = 2.26 \pm 0.56$. 
\label{fig4}} 
\end{figure}

We also searched for companion continuum emitters in these three HSC quasar fields. 
Here we conservatively define an emitter as one that shows $\geq 5\sigma$ significance. 
In the FoV of J1208$-$0200 ($5\sigma = 82$ $\mu$Jy beam$^{-1}$) 
and J2239+0207 ($5\sigma = 96$ $\mu$Jy beam$^{-1}$), no such emitter was found. 
In contrast, one emitter was found slightly outside the nominal FoV of J2228+0152 ($5\sigma = 56$ $\mu$Jy beam$^{-1}$), 
located 13\arcsec.7 away from the quasar position (Figure \ref{fig4}). 
This emitter is bright with a peak flux density of 0.25 mJy beam$^{-1}$, 
and is clearly more extended than the synthesized beam. 
The deconvolved size by simply fitting a two-dimensional Gaussian with the \verb|imfit| task is 
(0\arcsec.417 $\pm$ 0\arcsec.078) $\times$ (0\arcsec.306 $\pm$ 0\arcsec.080). 
We found that this object is also detected in the HSC optical bands as 
$g = 25.12 \pm 0.21$ mag, 
$r = 24.57 \pm 0.18$ mag, 
$i = 25.04 \pm 0.37$ mag, 
$z = 23.91 \pm 0.25$ mag, and $y = 24.32 \pm 0.82$, respectively. 
The HSC photometric redshift catalog \citep{2018PASJ...70S...9T} from the first data release \citep{2018PASJ...70S...8A}, 
along with the Bayesian-based Mizuki code \citep{2015ApJ...801...20T}, 
suggests that the redshift of this source is $z_{\rm photo} = 2.26 \pm 0.56$. 
If we rely on this $z_{\rm photo}$, the above source size is equivalent to $\sim 3.4$ kpc $\times$ 2.5 kpc. 
Such faint (e.g., a few 100 $\mu$Jy at $\sim 1$ mm) continuum sources have been uncovered 
by recent unbiased surveys of, e.g., HUDF/GOODS-S using ALMA 
\citep[e.g.,][]{2016ApJ...833...68A,2017MNRAS.466..861D,2018PASJ...70..105H,2018A&A...620A.152F}.

Combining with Cycle 4 data \citep{2018PASJ...70...36I}, we have observed seven HSC quasars in total with ALMA Band 6, 
and this object is the only one identified as a continuum emitter in the field, except for the target quasars. 
This detection rate (one continuum emitter in $7 \times 0.135$ arcmin$^2$ fields) 
seems to be smaller than recent measurements of the 1.2 mm number counts in the field 
\citep[e.g.,][]{2016ApJ...833...68A,2016ApJS..222....1F}. 
Indeed, the best-fit cumulative number count in \citet{2016ApJS..222....1F}
\footnote{We adopted $S_* = 2.35$ mJy, $\phi_* = 1.54 \times 10^3$ deg$^{-2}$, $\alpha = -2.12$, and effective area = 80\% of each field \citep[see Table 5 of][]{2016ApJS..222....1F}.} 
predicts that we should have detected $\sim 11-16$ continuum sources (when corrected for errors) over our seven fields, 
or $\sim 1-3$ in each 0.135 arcmin$^2$ field, given the depth of each one. 
This may be an overestimate for our fields because our high resolution observations are 
less sensitive to extended emission \citep[see discussion in][]{2017ApJ...850...83F}. 
We may also be subject to strong cosmic variance. 
Even so, our results will not support that these HSC quasars reside in overdense regions of submm sources. 
In this context, \citet{2018ApJ...867..153C} reported no submm overdensity 
around a sample of 35 $z > 6$ optically-luminous quasars (total effective area = 4.3 arcmin$^2$), 
although their observations are considerably shallower than ours. 
It is noteworthy, however, that some studies at $z \lesssim 5$ suggest that luminous quasars 
tend to reside in overdense regions of emitters \citep{2015ApJ...806L..25S,2017ApJ...836....8T}. 
Further observations are required to reconcile this discrepancy.

\section{Discussion}\label{sec4}
\subsection{The [C\,\emissiontype{II}]-FIR luminosity relation}\label{sec4.1}
\begin{figure}[h]
\begin{center}
\includegraphics[width=\linewidth]{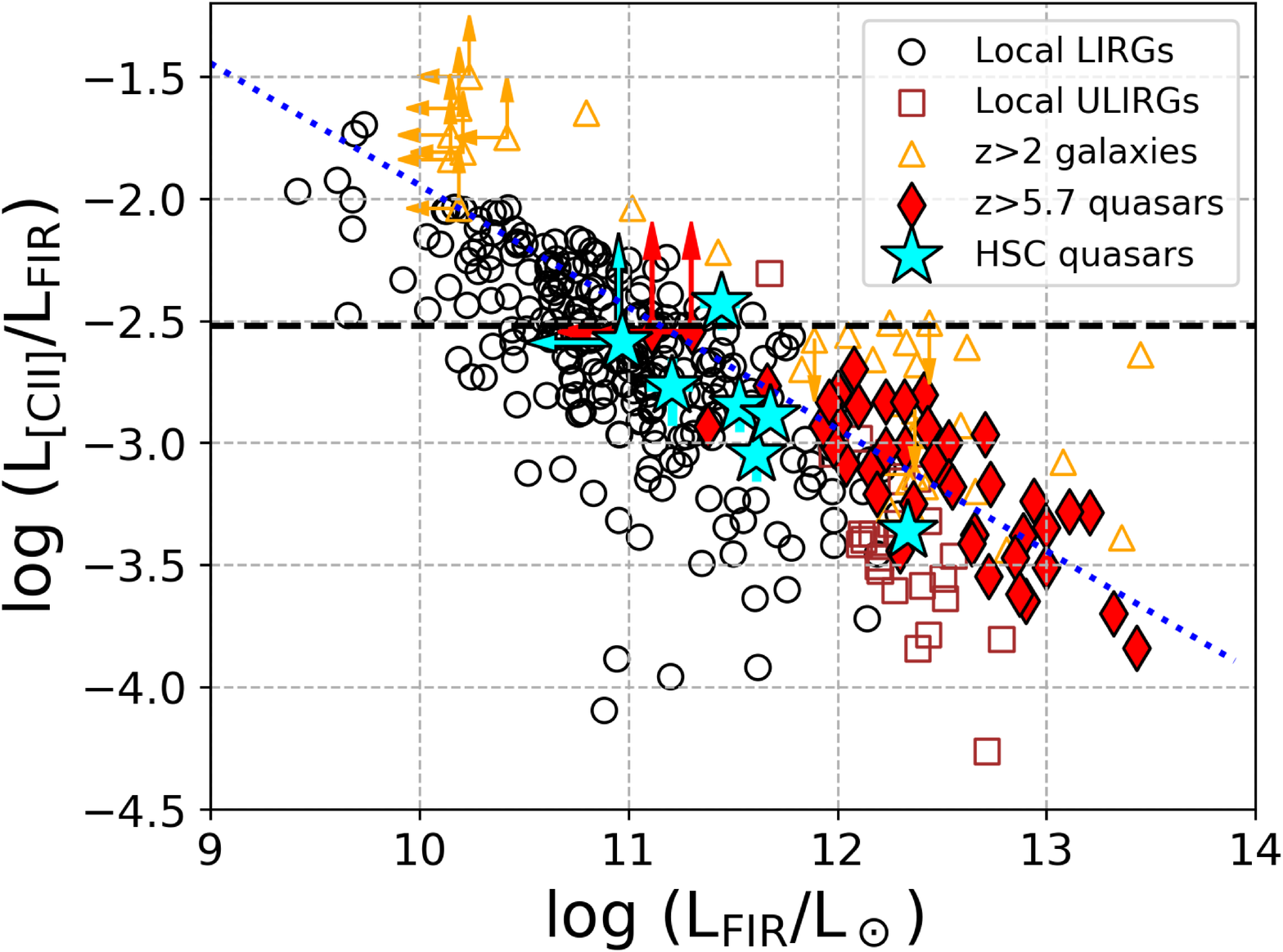}
\end{center}
\caption{
\cii to FIR luminosity ratio as a function of FIR luminosity for our HSC quasars 
(cyan stars; both the Cycle 4 and 5 samples are included): 
this is an updated version from the figure presented in \citep{2018PASJ...70...36I}. 
Compilations of various kinds of galaxies from literature are also plotted: 
local LIRGs \citep{2013ApJ...774...68D}, 
local ULIRGs \citep{2013ApJ...776...38F}, 
$z > 2$ FIR- or UV-luminous galaxies \citep{2009A&A...500L...1M,2010A&A...518L..35I,2011A&A...530L...8D,
2012ApJ...752L..30W,2013Natur.496..329R,2015MNRAS.449.2883G,2015Natur.522..455C}, 
and $z \gtrsim 6$ quasars \citep{2005A&A...440L..51M,2013ApJ...773...44W,2016ApJ...830...53W,
2015ApJ...805L...8B,2017ApJ...845..138S,2013ApJ...770...13W,2015ApJ...801..123W,2017ApJ...850..108W,2012ApJ...751L..25V,
2016ApJ...816...37V,2017ApJ...837..146V,2017ApJ...851L...8V,2018ApJ...866..159V, 
2017Natur.545..457D,2018ApJ...854...97D,2017ApJ...849...91M}. 
For all quasar samples, we assumed a gray body spectrum with $T_d = 47$ K and $\beta = 1.6$ 
to calculate $L_{\rm FIR}$ to maintain consistency. 
The horizontal dashed line indicates the Milky Way value \citep[$\sim 3 \times 10^{-3}$,][]{2013ARA&A..51..105C}. 
Where necessary, TIR luminosity was converted to FIR luminosity using $L_{\rm TIR} \simeq 1.3 L_{\rm FIR}$ \citep{2013ARA&A..51..105C}. 
The diagonal dotted line indicates our best-fit to the quasar data, excluding objects with upper and/or lower limits. 
}
\label{fig5} 
\end{figure}

We discuss the \cii to FIR luminosity ratio of our quasars and various comparison samples here. 
The ratio quantifies the contribution of \cii line emission to the cooling of the cold ISM 
\citep[the Milky Way value of $L_{\rm \cii}/L_{\rm FIR}$ is $3 \times 10^{-3}$,][]{2013ARA&A..51..105C}, 
but it has long been known that the $L_{\rm \cii}/L_{\rm FIR}$ ratio is an order of magnitude smaller in ULIRG-like FIR-luminous systems 
\citep[e.g.,][]{1997ApJ...491L..27M,2008ApJS..178..280B,2010ApJ...724..957S,2011ApJ...728L...7G,2013ApJ...776...38F,2013ApJ...774...68D,2017ApJ...846...32D}. 
This {\it [C\,\emissiontype{II}]-deficit} trend has also been found in high-redshift quasars 
\citep[e.g.,][]{2013ApJ...773...44W,2016ApJ...816...37V,2018ApJ...854...97D}. 
Several processes may contribute to the deficit in quasars, including 
an AGN contribution to $L_{\rm FIR}$ \citep{2014ApJ...790...15S}, 
reduction of C$^+$ abundance due to AGN irradiation \citep{2015A&A...580A...5L}, 
high gas surface densities of individual clouds \citep[giving a high molecular-to-atomic gas fraction,][]{2017MNRAS.467...50N}, etc., 
but an overwhelming factor appears to be the presence of a high FIR luminosity density region and/or high-temperature dust-emitting region, 
as has been invoked for local ULIRGs \citep[e.g.,][see also the discussion in the last part of this subsection]{1997ApJ...491L..27M,2013ApJ...774...68D,2017ApJ...846...32D}.

\begin{figure}[h]
\begin{center}
\includegraphics[width=\linewidth]{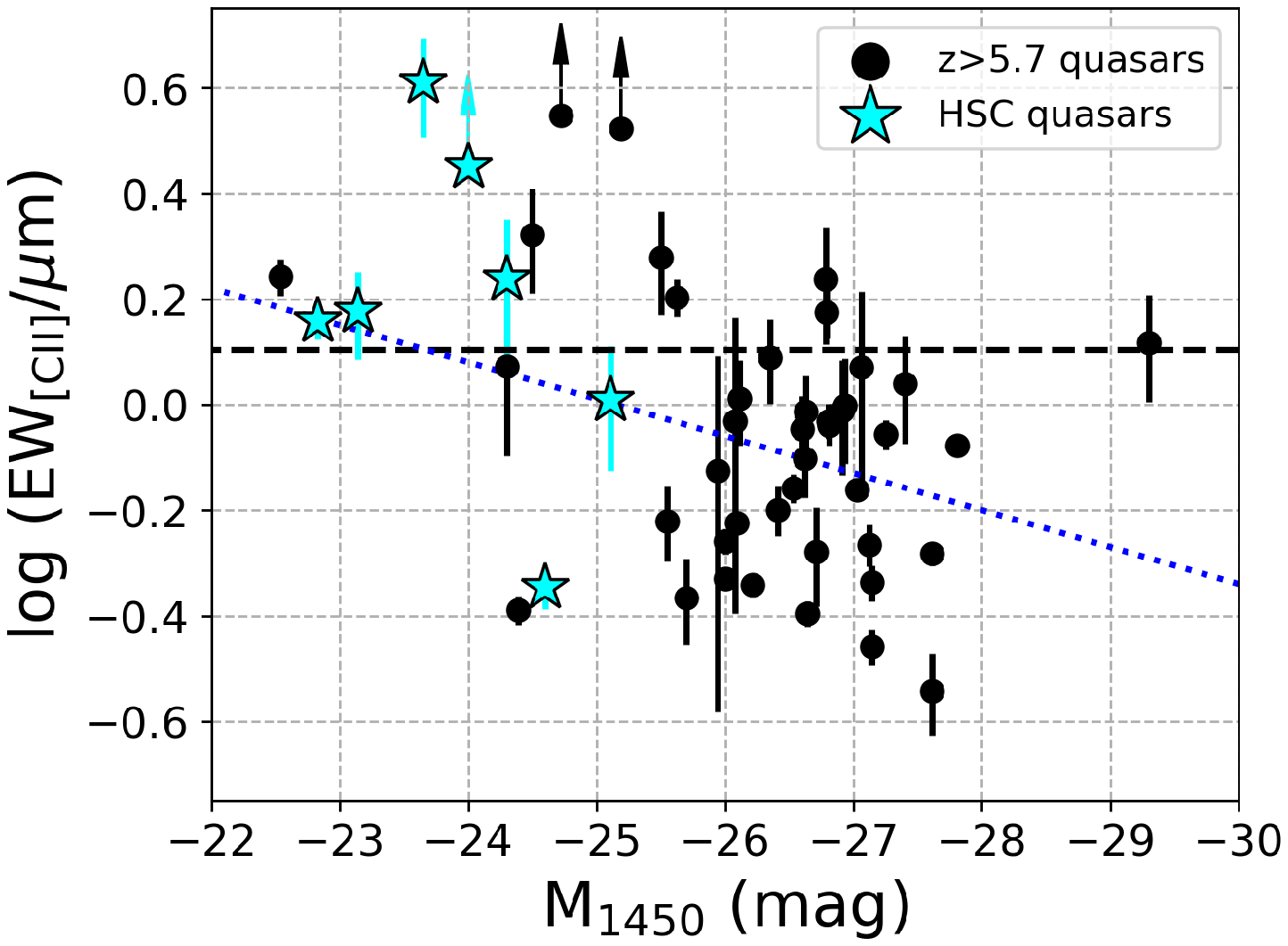}
\end{center}
\caption{
Rest frame \cii line equivalent width (EW$_{\rm \cii}$) as a function of the quasar absolute UV magnitude ($M_{\rm 1450}$), 
using the same quasar samples as shown in Figure \ref{fig5}. 
The mean EW for local starburst galaxies is indicated by the horizontal dashed line \citep{2013ApJ...774...68D}. 
Our best power-law fit (excluding objects with lower limits) is shown by the diagonal dotted line. 
\label{fig6}} 
\end{figure}

Figure \ref{fig5} shows the [C\,\emissiontype{II}]-deficit trend with a compilation 
of galaxies at various redshifts, including $z > 5.7$ quasars with available data 
(the 7 HSC quasars + 43 previously-studied objects). 
By adding optically-selected lower-luminosity 
($L_{\rm FIR} \sim 10^{11}~L_\odot$) objects like our HSC quasars, 
we have increased the dynamic range of the plot, making it easier 
to see any correlation between $L_{\rm FIR}$ and $L_{\rm \cii}/L_{\rm FIR}$. 
Here the $L_{\rm \cii}/L_{\rm FIR}$ ratios of the HSC quasars are not drastically different from 
those of low redshift galaxies with comparable $L_{\rm FIR}$, except for the ULIRG-class object J2239+0207. 
The modest ratios found for the HSC quasars were also observed in comparably 
optically- and FIR-faint CFHQS quasars \citep{2013ApJ...770...13W,2015ApJ...801..123W,2017ApJ...850..108W}. 
Note that some of the previously-studied quasars may be biased 
toward low $L_{\rm \cii}/L_{\rm FIR}$ values as they were originally selected based on their high $L_{\rm FIR}$. 
However, as recent ALMA follow-up studies have been performed for 
a number of optically-selected quasars \citep[e.g.,][]{2018ApJ...854...97D}, 
the trend in Figure \ref{fig5} becomes less biased. 
The Spearman rank correlation coefficient for the whole quasar sample in Figure \ref{fig5}, 
excluding objects with upper or lower limits, is fairly high ($\rho = -0.78$) 
with a null-hypothesis probability of $p = 1.29 \times 10^{-10}$. 
The relationship can be expressed as 
\begin{equation}
\log \left( \frac{L_{\rm [C\,\emissiontype{II}]}}{L_{\rm FIR}} \right) = (3.05 \pm 0.67) + (-0.50 \pm 0.05) ~\log \left( \frac{L_{\rm FIR}}{L_\odot} \right), 
\end{equation}
which is consistent with previous analysis \citep{2017ApJ...850..108W,2018PASJ...70...36I}.

\begin{figure}[h]
\begin{center}
\includegraphics[width=\linewidth]{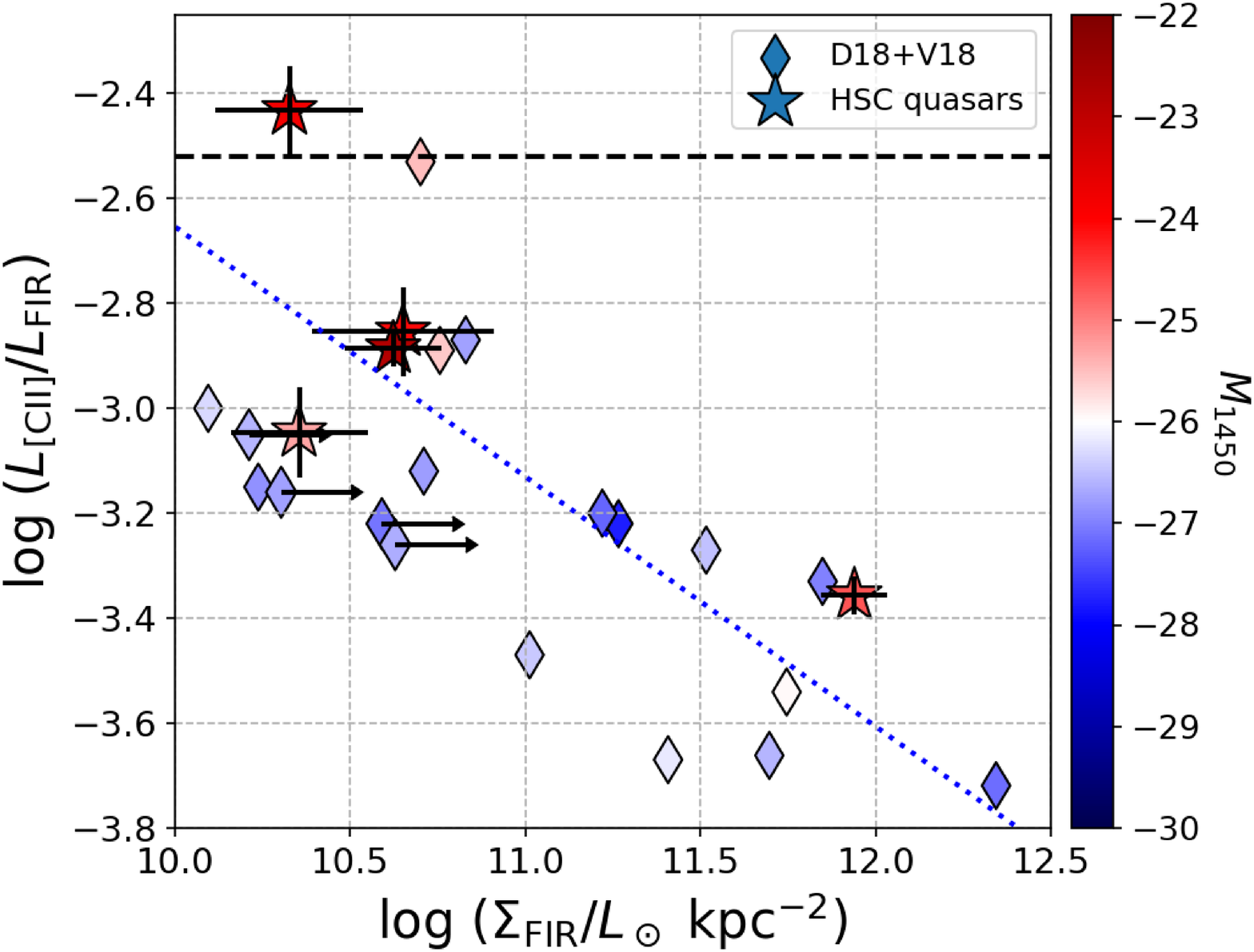}
\end{center}
\caption{
\cii to FIR luminosity ratio as a function of FIR luminosity density 
($\Sigma_{\rm FIR} = L_{\rm FIR}/2\pi R^2_{\rm FIR}$, $R_{\rm FIR} = {\rm FWHM}_{\rm maj}/2$) on a logarithmic scale, 
for the HSC quasars and quasars at $z > 6$ compiled from \citet{2018ApJ...854...97D} and \citet{2018ApJ...866..159V}. 
We assumed the gray body spectrum with $T_d = 47$ K and $\beta = 1.6$ for all quasars to compute $L_{\rm FIR}$. 
The quasars are color-coded by their $M_{\rm 1450}$ values. 
Only objects with reliable measurements of $R_{\rm FIR}$ are included. 
The horizontal dashed line denotes the Milky Way value \citep{2013ARA&A..51..105C} for an eye guide. 
Our best power-law fit (excluding objects with lower limits on $\Sigma_{\rm FIR}$) is indicated by the blue dotted line. 
\label{fig7}} 
\end{figure}

In order to investigate the physical origin of the [C\,\emissiontype{II}]-deficit in high-redshift quasars, 
we measured their rest frame \cii equivalent width (EW$_{\rm \cii}$) to explore the quasar contribution to the FIR light, defined as 
\begin{equation}
\frac{{\rm EW}_{\rm \cii}}{\rm \mu m} = 0.527 \times \frac{S_{\rm \cii}~{\rm [Jy~km~s^{-1}]}}{f_{\rm cont}~{\rm [mJy]}}, 
\end{equation} 
where the continuum emission is measured at frequencies close to that of the line. 
The use of EW$_{\rm \cii}$ has the advantage that it does not require any assumption about the shape of the IR SED. 
With quasar samples in Figure \ref{fig5}, we found a correlation between 
EW$_{\rm \cii}$ and the quasar absolute UV magnitude (Figure \ref{fig6}) as 
\begin{equation}
\log \left( \frac{\rm EW_{\rm \cii}}{\rm \mu m} \right) = (1.76 \pm 0.64) + (0.07 \pm 0.02) ~\left( \frac{M_{\rm 1450}}{\rm mag} \right). 
\end{equation}
However, this correlation is marginal with $\rho = 0.32$ and $p = 0.03$; 
i.e., its significance is much weaker than the $L_{\rm FIR}$--$L_{\rm \cii}/L_{\rm FIR}$ correlation. 
Indeed, there are a number of optically-luminous quasars that show comparable 
EW$_{\rm \cii}$ to the much fainter HSC quasars (Figure \ref{fig6}). 
Thus, while the marginal correlation coefficient (0.32) implies a certain level of 
quasar contribution to the 1.2 mm continuum flux density (hence $L_{\rm FIR}$), 
it is not a prime driver of the [C\,\emissiontype{II}]-deficit. 
Furthermore, given the positive correlation observed between quasar nuclear luminosity 
and $L_{\rm \cii}$ (Figure \ref{fig_rev2}), 
AGN irradiation does not play a primary role in the [C\,\emissiontype{II}]-deficit. 

Another plausible factor is the high FIR luminosity density and/or 
the existence of a high temperature dust-emitting region \citep[e.g.,][]{2013ApJ...774...68D,2017ApJ...846...32D,2017ApJ...834....5S,2018ApJ...861...95H}. 
In this context, Figure \ref{fig7} investigates the dependence of the $L_{\rm \cii}/L_{\rm FIR}$ ratio 
on the FIR luminosity density ($\Sigma_{\rm FIR} = L_{\rm FIR}/2\pi R^2_{\rm FIR}$), 
computed using the two-dimensional Gaussian fit size in the FIR continuum map. 
In addition to the HSC quasars, we plot $z \gtrsim 6$ quasars compiled from 
\citet{2018ApJ...854...97D} and \citet{2018ApJ...866..159V}. 
These quasars were selected from optical surveys, and thus are free from a $L_{\rm FIR}$-based selection bias. 
We then found a strong anti-correlation between $L_{\rm \cii}/L_{\rm FIR}$ and $\Sigma_{\rm FIR}$: 
\begin{equation}
\log \left( \frac{L_{\rm [C\,\emissiontype{II}]}}{L_{\rm FIR}} \right) = (2.11 \pm 1.00) + (-0.48 \pm 0.09) ~\log \left( \frac{\Sigma_{\rm FIR}}{L_\odot~{\rm kpc^2}} \right), 
\end{equation} 
with $\rho = -0.78$ and $p = 4.9 \times 10^{-5}$, respectively. 
This strong trend is consistent with previous results for $z \gtrsim 6$ quasars \citep{2018ApJ...854...97D}. 
It is noteworthy that the local ULIRGs also fit to this scenario well, as they typically 
have even smaller FIR emitting regions \citep[$\lesssim 500$ pc, e.g.,][]{2000AJ....119..509S,2008ApJ...684..957S,2011AJ....141..156I} 
than high-$z$ submillimeter galaxies and quasar hosts having comparable $L_{\rm FIR}$ 
\citep[e.g.,][]{2016ApJ...833..103H,2017ApJ...850...83F}, 
which will lead to the their small $L_{\rm \cii}/L_{\rm FIR}$ ratios seen in Figure \ref{fig5}. 
Note that recent very high resolution ALMA observations enabled spatially-resolved 
measurements of $L_{\rm \cii}/L_{\rm FIR}$, which revealed similarly small ratios 
in the central ($r \lesssim 1$ kpc) high $\Sigma_{\rm FIR}$ regions of some submillimeter galaxies \citep[SMGs, e.g.,][]{2018ApJ...859...12G,2019ApJ...876..112R}.

There are three intimately linked physical processes which may drive the [C\,\emissiontype{II}]-deficit in high $\Sigma_{\rm FIR}$ regions. 
First process is driven by the increased radiation field, 
under which dust particles have more positive charge. 
This results in a reduction of the number of free electrons 
released from the dust particles, which contribute to [C\,\emissiontype{II}] excitation 
\citep[e.g.,][]{1997ApJ...491L..27M,2001A&A...375..566N}. 
Second, increased ionized-to-atomic hydrogen ratio will reduce the fraction of UV photons absorbed by gas, 
which then leads to the reduction of the $L_{\rm \cii}/L_{\rm FIR}$ ratio \citep{2009ApJ...701.1147A}. 
Third, the temperature of the dust ($T_{\rm d}$) itself matters. 
Dust grains are heated to higher $T_{\rm d}$ in higher $\Sigma_{\rm FIR}$ regions, 
in which a larger number of ionizing photons is available, as seen in local ULIRGs. 
This greatly enhances $L_{\rm FIR}$ and thus reduces $L_{\rm \cii}/L_{\rm FIR}$ 
\citep[e.g.,][]{2013ApJ...774...68D,2017ApJ...846...32D}. 
To further test this possibility, however, shorter wavelength continuum 
observations are necessary to constrain $T_{\rm d}$ directly 
as we now assumed $T_{\rm d} = 47$ K \citep{2006ApJ...642..694B} 
for all quasars in Figure \ref{fig7} for consistency.

\subsection{Early SMBH--host galaxy co-evolution}\label{sec4.2}
We now investigate the early co-evolution of 
SMBHs and their host galaxies, in both integrated and differential forms. 
The relevant properties of the seven HSC quasars with ALMA data 
are compared with those of previously observed $z \gtrsim 6$ quasars, 
to give a less-biased view on early mass assembly. 
We recall that the HSC quasars constitute the break ($M^\star_{\rm 1450} = -24.9$ mag) 
or further lower luminosity regime of the quasar luminosity function at $z \sim 6$ \citep{2018ApJ...869..150M}, 
which then represent the bulk of the quasar population at that era. 
Subsequent NIR follow-up observations started to reveal that they 
possess a wide range of $M_{\rm BH}$ \citep[$\sim 10^{7.5}-10^{9}~M_\odot$,][]{Onoue19}, 
which are therefore characterized by a wide range of Eddington ratio ($\sim 0.1-1$). 
Among the six HSC quasars reported in \citet{Onoue19}, 
four objects indeed have Eddington ratios of $0.16 - 0.24$. 
Note that dust obscuration does not play a major role in 
shaping the low-luminosity nature of most of the HSC quasars, 
as judged from the rest-UV SED modeling \citep{Onoue19}. 

\subsubsection{Integrated form: $M_{\rm BH} - M_{\rm dyn}$}\label{sec4.2.1}
\begin{table*}[h]
  \tbl{Dynamical properties of the HSC quasars observed in ALMA Cycle 5}
  {%
  \begin{tabular}{cccc} \hline \hline
   Name & $M_{\rm dyn} \sin^2i$ (10$^{10}$ $M_\odot$) & $M_{\rm dyn}$ (10$^{10}$ $M_\odot$) & $M_{\rm BH}$ (10$^8$ $M_\odot$) \\ \hline
   J1208$-$0200 & 1.2 $\pm$ 0.3 & 1.3 & 7.1$^{+2.4}_{-5.2}$ \\
   J2228+0152 & 1.5 $\pm$ 0.4 & 2.0 & $>$1.1 \\
   J2239+0207 & 6.4 $\pm$ 1.3 & 29 & 11$^{+3}_{-2}$ \\ \hline
  \end{tabular}}\label{tbl4}
  \begin{tabnote}
    Formal errors on $M_{\rm dyn}$ are not given due to multiple unconstrained uncertainties 
    including those of the inclination angles and the geometry of the line emitting regions. 
    $M_{\rm BH}$ of J1208$-$0200 and J2239+0207 are measured with Mg\,\emissiontype{II} emission line, which are reported in \citet{Onoue19}. 
    Meanwhile, Eddington-limited mass accretion is assumed for J2228+0152, 
    giving the lower limit on its $M_{\rm BH}$. 
    By following previous works, we assume a typical systematic uncertainty for 
    the Mg\,\emissiontype{II}-based $M_{\rm BH}$ of 0.5 dex. 
  \end{tabnote}
  \end{table*}

We first computed $M_{\rm BH}$ of J1208$-$0200 and J2239+0207 (Table \ref{tbl4}) 
using the broad Mg\,\emissiontype{II}-based virial mass calibration 
\citep[the so-called single epoch method,][]{2009ApJ...699..800V}. 
Details of the procedure are described in \citet{Onoue19}. 
J2228+0152 has no Mg\,\emissiontype{II} spectroscopy, so we assumed 
Eddington-limited accretion to derive its $M_{\rm BH}$. 
Note that this assumption is often made in $z \gtrsim 6$ quasar studies 
\citep[e.g.,][]{2013ApJ...773...44W,2018ApJ...854...97D} to compute the {\it lower limit} of $M_{\rm BH}$. 
The bolometric luminosity was calculated from the 1450 {\AA} monochromatic luminosity 
with a correction factor of 4.4 \citep{2006ApJS..166..470R}. 
The estimated $M_{\rm BH}$ of the three HSC quasars presented here fall in the range $(1.1 - 11) \times 10^8~M_\odot$. 
Along with our Cycle 4 measurements, these HSC quasars populate 
the middle to lower regime of the $z \gtrsim 6$ quasar mass distribution observed thus far \citep{Onoue19}. 
Hence we are indeed probing a quasar population less biased in terms of $M_{\rm BH}$ 
by observing these low-luminosity quasars. 

Next, we estimate their host galaxy dynamical masses ($M_{\rm dyn}$) 
from the \cii spatial extents and line widths, by following the standard procedure used in $z \gtrsim 6$ quasar studies 
\citep{2013ApJ...773...44W,2015ApJ...801..123W,2017ApJ...850..108W,2016ApJ...816...37V,2018PASJ...70...36I}. 
Here the \cii emission is assumed to originate from a thin rotating circular disk. 
The inclination angle ($i$) of the disk is determined from the axis ratio 
of the deconvolved Gaussian fit to the \cii emitting region. 
The circular velocity is expressed as $v_{\rm circ} = 0.75 {\rm FWHM}_{\rm \cii}/\sin i$. 
The disk size is given by $D = 1.5 \times a_{\rm maj}$, where $a_{\rm maj}$ is the deconvolved size 
of the Gaussian major axis, and the factor 1.5 is used to account for 
spatially extended low level emission \citep[e.g.,][]{2010ApJ...714..699W}. 
The $M_{\rm dyn}$ within $D$ is then, 
\begin{equation}
M_{\rm dyn}/M_\odot = 1.16 \times 10^5 \left( \frac{v_{\rm circ}}{{\rm km~s^{-1}}} \right)^2 \left( \frac{D}{\rm kpc} \right)
\end{equation}
The resultant values for both $M_{\rm dyn} \sin^2 i$ and $M_{\rm dyn}$ are listed in Table \ref{tbl4}. 
Note that formal errors on $M_{\rm dyn}$ are not given due to multiple unconstrained uncertainties 
including the inclination angles and the true geometry of the \cii emitting regions. 
Moreover, if the host galaxies have dispersion-dominant gas dynamics, 
their $M_{\rm dyn}$ will be significantly smaller than the ones 
derived with the rotating disk assumption \citep[e.g.,][]{2017ApJ...837..146V}, 
which will affect our discussion in the following. 
Unfortunately, it is difficult at this moment to investigate if the host galaxies discussed here 
are indeed rotation-dominant systems, as the spatial resolutions obtained thus far are only modest in most cases. 
Further higher resolution ALMA observations will reveal the dynamical nature of $z \gtrsim 6$ quasars 
\citep[see recent examples in,][]{2019ApJ...874L..30V,2019arXiv190606801W}, 
and give a better insight on the early co-evolution. 
Despite these large uncertainties, we hereafter use $M_{\rm dyn}$ 
as a surrogate for the stellar mass ($M_\star$) of the quasar hosts, 
as is often done in high redshift quasar studies 
\citep[e.g.,][]{2013ApJ...773...44W,2015ApJ...801..123W,2016ApJ...816...37V,2018PASJ...70...36I}. 
The derived values of $M_{\rm dyn}$ exceed $10^{10}~M_\odot$, or even $10^{11}~M_\odot$ for the case of J2239+0207, 
which lie at the massive end of the $M_\star$-distribution for $z \sim 6$ galaxies in general \citep[e.g.,][]{2015A&A...575A..96G}. 
Thus, the host galaxies of the HSC quasars are among the most evolved systems known at $z \sim 6$ in terms of their galaxy-masses. 

We also compiled $M_{\rm BH}$ and $M_{\rm 1450}$ values of other $z \geq 5.7$ quasars 
from the literature \citep{2010AJ....140..546W,2014ApJ...790..145D,2015ApJ...798...28K,
2015ApJ...801L..11V,2016ApJS..227...11B,2016ApJ...833..222J,2017ApJ...845..138S,2017ApJ...849...91M,2018ApJ...854...97D}. 
Most of the $M_{\rm BH}$ values from the  literature were indeed measured with the same 
Mg\,\emissiontype{II}-based single epoch method that we have used above \citep{2009ApJ...699..800V}. 
Exceptions are J1319+0950 \citep{2017ApJ...845..138S}, 
as well as J0842+1218, J1207+0630, and J2310+1855 \citep{2019ApJ...873...35S}, 
for which we recalculated $M_{\rm BH}$ with the \citet{2009ApJ...699..800V} calibration. 
For those without available $M_{\rm BH}$ data, we assumed the Eddington-limited accretion. 
Although this assumption may not be valid for our HSC quasar J2228+0152 
as it shows a rather quiescent nature (non-detection of FIR continuum emission), 
we nevertheless keep this assumption as we would like to take a statistical approach. 
The $M_{\rm dyn}$ of these quasars from the literature were also 
computed in the same way we adopted for our HSC quasars, 
i.e., the thin disk approximation with a 2D Gaussian decomposition \citep{2013ApJ...773...44W,2013ApJ...770...13W,2015ApJ...801..123W,
2017ApJ...850..108W,2016ApJ...816...37V,2017ApJ...845..138S,2018ApJ...854...97D}. 
Quasars that do not have such a Gaussian decomposition were excluded from our sample. 
One exception is the second highest redshift quasar known, J1120+0641 \citep[$z = 7.08$,][]{2011Natur.474..616M}. 
Previous high resolution \cii observations revealed that its host galaxy does not show ordered rotation, 
and an upper limit on $M_{\rm dyn}$ was provided by applying the virial theorem \citep{2017ApJ...837..146V}.

\begin{figure*}
\begin{center}
\includegraphics[width=\linewidth]{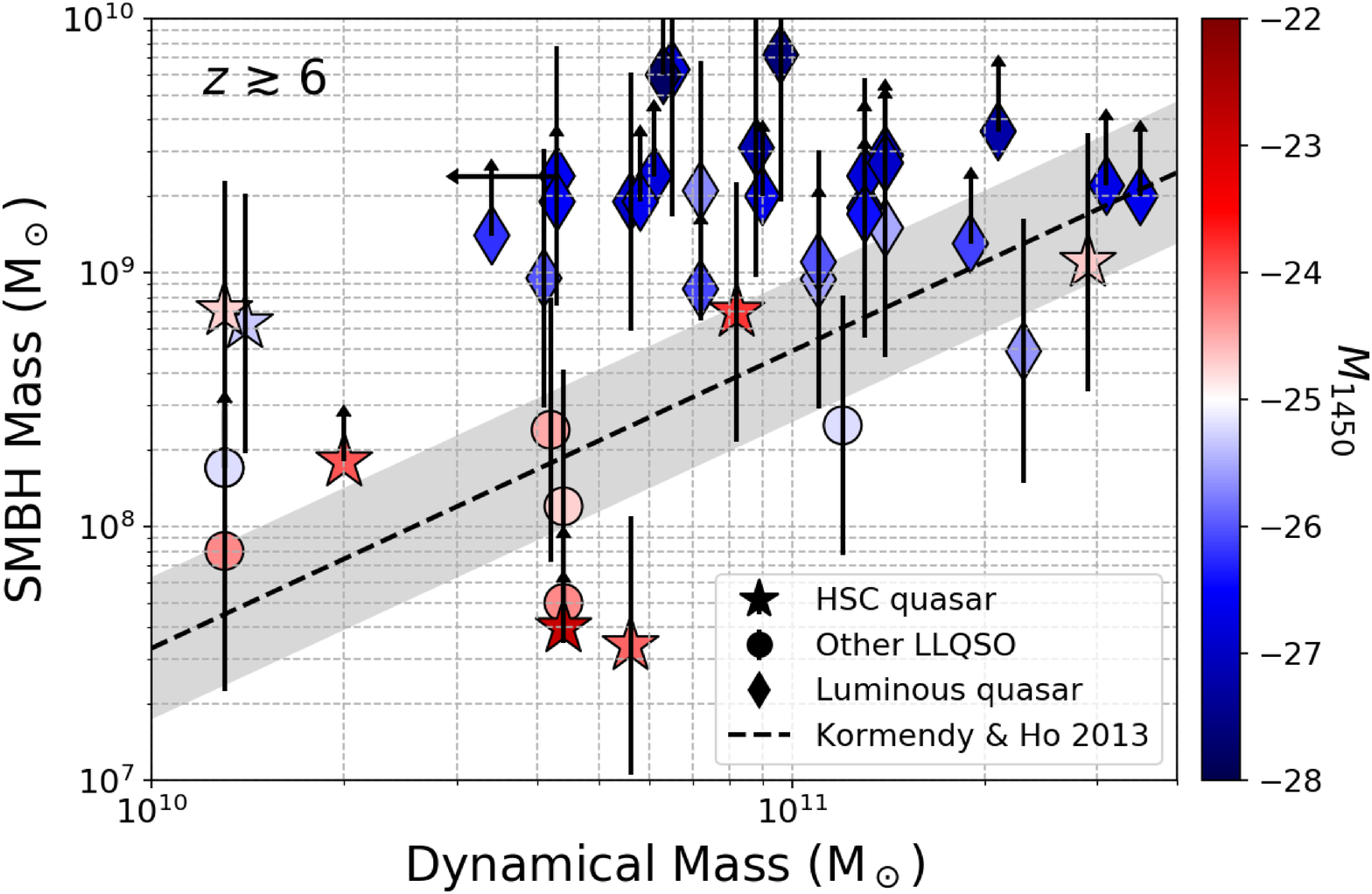}
\end{center}
\caption{
Black hole mass ($M_{\rm BH}$) vs host galaxy dynamical mass ($M_{\rm dyn}$) relationship for $z \gtrsim 6$ quasars, 
color-coded by their absolute UV magnitude ($M_{\rm 1450}$). 
The diagonal dashed line and the shaded region indicate the local $M_{\rm BH}$--$M_{\rm bulge}$ relationship 
and its 1$\sigma$ scatter, respectively \citep{2013ARA&A..51..511K}: 
we equate $M_{\rm  dyn}$ and $M_{\rm bulge}$ in this plot. 
From this figure, it is clear that optically luminous quasars ($M_{\rm 1450} \lesssim -25$ mag) 
typically show overmassive $M_{\rm BH}$ with respect to the local relation. 
On the other hand, a large fraction of the low-luminosity ($M_{\rm 1450} \gtrsim -25$ mag) quasars, 
including most of the HSC quasars, lie close to, or even below, that relation. 
See the main text for the samples of objects shown. 
\label{fig8}} 
\end{figure*}

In Figure \ref{fig8} we display the relation between $M_{\rm BH}$ and $M_{\rm dyn}$ for the 40 above-mentioned quasars at $z \gtrsim 6$, 
overlaid with the local $M_{\rm BH} - M_{\rm bulge}$ relation 
after equating $M_{\rm dyn}$ to $M_{\rm bulge}$ \citep{2013ARA&A..51..511K}. 
These quasars are color-coded by their $M_{\rm 1450}$. 
Regarding the optically-luminous objects, this figure supports conclusions in previous works 
\citep[e.g.,][]{2013ApJ...778..179W,2016ApJ...816...37V}. 
That is, the luminous quasars ($M_{\rm 1450} \lesssim -25$ mag) typically have overmassive SMBHs 
with respect to the local relation, while the discrepancy becomes less evident at $M_{\rm dyn} \gtrsim 10^{11}~M_\odot$: 
some luminous quasars even show comparable $M_{\rm BH}/M_{\rm dyn}$ ratios to the local relation. 
It is likely for such high-mass host galaxies that past multiple major and minor mergers already led to this convergence. 

On the other hand, most of the low-luminosity quasars with $M_{\rm 1450} \gtrsim -25$ mag 
show comparable ratios to, or even lower ratios than, the local relation, 
particularly at a range of $M_{\rm dyn} \gtrsim 4 \times 10^{10}~M_\odot$. 
The existence of the undermassive SMBHs even implies an evolutionary path, 
in which galaxies grow earlier than SMBHs, such as expected in a standard 
merger-induced evolution model \citep[e.g.,][]{2008ApJS..175..356H}. 
In this high $M_{\rm dyn}$ range, our result demonstrates 
that previous works on luminous quasars have been largely biased 
toward the most massive SMBHs \citep{2007ApJ...670..249L,2014MNRAS.438.3422S}, 
easily resulting in objects lying above the local $M_{\rm BH} - M_{\rm dyn}$ relation. 
Therefore, our result highlights the power of the sensitive Subaru survey 
to probe the fainter part of the quasar luminosity- and mass-functions 
and to reveal the nature of early co-evolution of black holes and their host galaxies in a less biased way. 

However, we also found that the low-luminosity quasars at $M_{\rm dyn} \lesssim 3 \times 10^{10}~M_\odot$ 
start to show overmassive $M_{\rm BH}$ with respect to the local relation. 
Regarding the HSC quasars, we argue that this is still due to our selection bias: 
we basically selected objects with $M_{\rm 1450} < -24$ for the targets in our NIR follow-up observations \citep{Onoue19}, 
which corresponds to $M_{\rm BH} \sim 10^8~M_\odot$ if it is accreting at the Eddington-limit. 
Meanwhile, the $M_{\rm BH}$ expected at $M_{\rm dyn} \sim (1-3) \times 10^{10}~M_\odot$ 
from the local relation is well smaller than $10^8~M_\odot$. 
Thus, given the large scatter of the Eddington ratio distribution at $z \sim 6$ \citep[$\sim 0.01-1$,][]{Onoue19}, 
we clearly need to observe much fainter objects to surely probe the $M_{\rm BH} < 10^8~M_\odot$ region and to reveal the unbiased shape of the relation. 
Future sensitive observations with the {\it James Webb Space Telescope (JWST)} 
or ground-based extremely large telescopes will allow to probe down this very low mass region even at $z > 6$. 

Given their low nuclear luminosities (hence mass accretion rate onto SMBHs) and SFRs, 
the low-luminosity quasars studied here will not move drastically in the $M_{\rm BH} - M_{\rm dyn}$ plane 
even over the next $\sim 10$ Myr \citep[typical life-time of high-redshift quasars expected with the transverse proximity effect,][]{2016ApJ...830..120B}, 
unless there is a rich supply of gas which infalls to the galaxy. 
This is particularly true if the quasar duty-cycle is $\lesssim 10^{-2}$ 
as suggested for $z \sim 6$ objects recently \citep{2018ApJ...868..126C}. 
Note that \citet{2018PASJ...70...36I} argued that the four HSC quasar hosts studied in ALMA Cycle 4 
are on or even below the star formation main sequence at $z \sim 6$ \citep[e.g.,][]{2015ApJ...799..183S}. 
This trend also holds for the Cycle 5 samples. 
Therefore, these low-luminosity quasars may already be near the end of their SMBH growth and galaxy growth, 
and are transforming into a quiescent phase even at $z \gtrsim 6$. 
This is consistent with the conclusion of \citet{Onoue19}, who raised a similar argument based on the Eddington ratio measurements for HSC quasars. 
If this is true, there must be a quite rapid physical process to realize the $M_{\rm BH} - M_{\rm dyn}$ relation at that high redshift, 
such as expected in a merger-induced evolution \citep[e.g.,][]{2005Natur.433..604D,2006ApJS..163....1H,2008ApJS..175..356H}. 
Indeed, a recent very high resolution simulation based on this scheme suggests that 
even a quasar at $z = 7$ follows the local co-evolution relation \citep{2019arXiv190102464L}.

\subsubsection{Differential form: $L_{\rm Bol} - L_{\rm FIR}$}\label{sec4.2.2}
\begin{figure}
\begin{center}
\includegraphics[width=\linewidth]{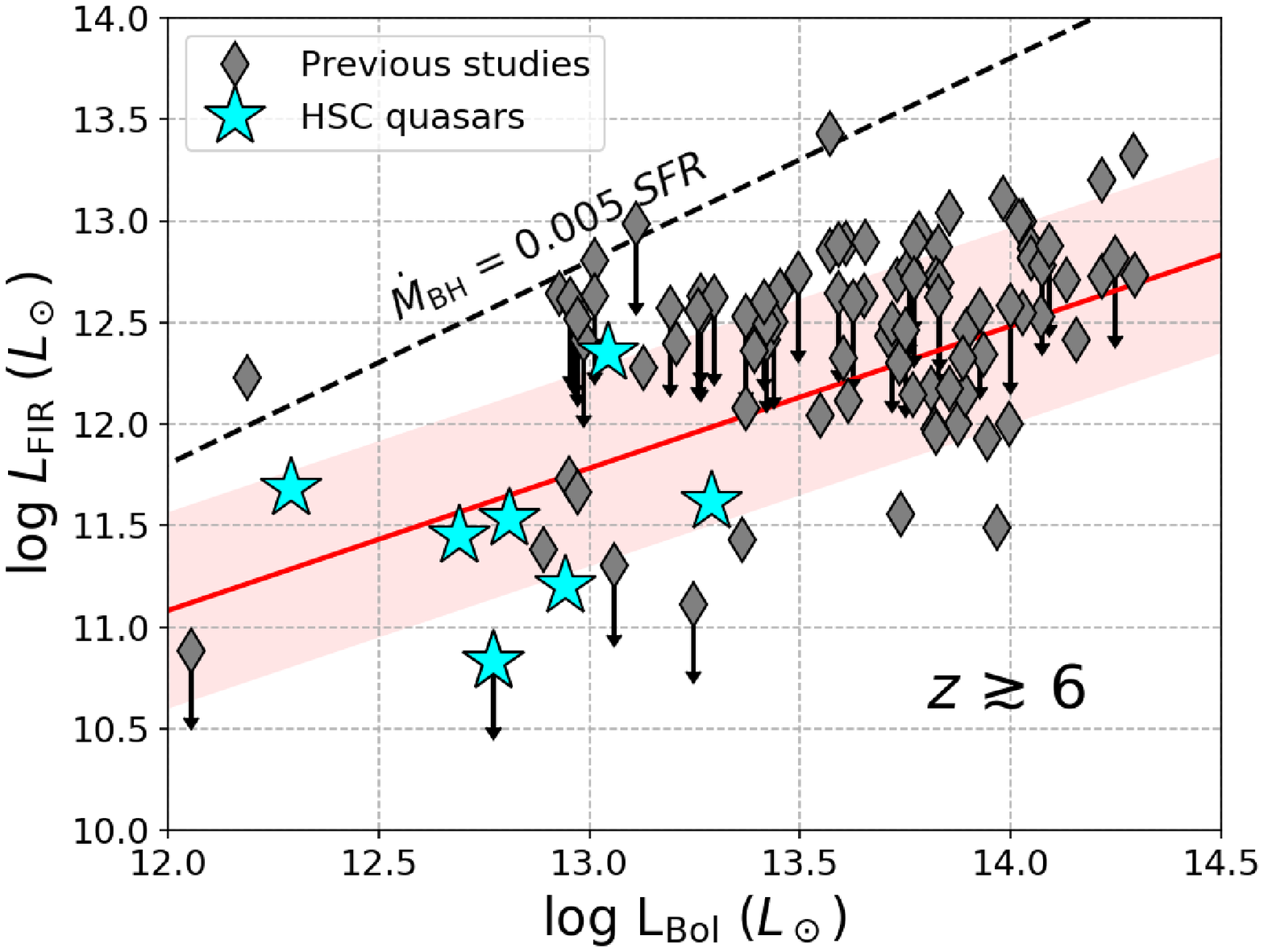}
\end{center}
\caption{
$L_{\rm Bol} - L_{\rm FIR}$ relationship for 97 $z \gtrsim 6$ quasars 
including the HSC quasars, on a logarithmic scale. 
The dotted line (red) and the shaded region indicate the best fit linear regression for the whole sample 
(including objects with upper limits on $L_{\rm FIR}$) and its 1$\sigma$ scatter, respectively. 
The dashed line (black) refers to the so-called {\it parallel growth model}, in which SMBHs and galaxies 
grow simultaneously by following the local $M_{\rm BH} - M_{\rm bulge}$ relation \citep{2013ARA&A..51..511K}. 
We define the quasar-dominant phase as the region below this parallel growth line, 
whereas the starburst-dominant phase is above the line. 
\label{fig9}} 
\end{figure}

The $M_{\rm BH} - M_{\rm dyn}$ relationship studied in \S~\ref{sec4.2.1} reflects 
the integrated history of past mass accumulation. 
We now investigate the on-going mass accumulation of $z \gtrsim 6$ quasars 
by comparing quasar bolometric luminosity ($L_{\rm Bol}$) and $L_{\rm FIR}$. 
As $L_{\rm Bol}$ and $L_{\rm FIR}$ can be converted to the growth rate 
of the central SMBH ($\equiv \dot{M}_{\rm BH}$) and the SFR of the host galaxy (see \S~\ref{sec3.2}), respectively, 
the $L_{\rm Bol} - L_{\rm FIR}$ relation indicates a {\it differential} form of the Maggorian relation \citep[e.g.,][]{2013ApJ...770...13W,2014A&A...566A..53D}. 

We again compiled literature data for $M_{\rm 1450}$ 
\citep{2003ApJ...587L..15W,2010AJ....140..546W,2014ApJ...790..145D,
2015ApJ...798...28K,2015Natur.518..512W,2016ApJ...833..222J,
2017ApJ...849...91M,2017ApJ...845..138S,2016ApJS..227...11B,
2018Natur.553..473B,2018ApJ...854...97D,2018ApJ...866..159V} 
and $L_{\rm FIR}$ \citep{2003A&A...406L..55B,2003AJ....126...15P,
2005A&A...440L..51M,2007AJ....134..617W,2008ApJ...687..848W,
2011AJ....142..101W,2013ApJ...773...44W,2016ApJ...830...53W,
2013A&A...552A..43O,2013ApJ...770...13W,2015ApJ...801..123W,
2017ApJ...850..108W,2015ApJ...805L...8B,2016ApJ...816...37V,
2017ApJ...845..138S,2017ApJ...837..146V,2017ApJ...851L...8V,
2018ApJ...866..159V,2017ApJ...849...91M,2017Natur.545..457D,2018ApJ...854...97D} 
of $z \gtrsim 6$ quasars. 
In addition to the objects we compiled in \S~\ref{sec4.2.1}, 
we appended a number of quasars, mostly with single dish $L_{\rm FIR}$ measurements. 
Thus the total number of quasars on this analysis is increased to 97, including the seven HSC quasars. 
Their $L_{\rm Bol}$ and $L_{\rm FIR}$ values were computed in the same manner as described earlier. 

Figure \ref{fig9} shows the $L_{\rm Bol} - L_{\rm FIR}$ relationship 
for the $z \gtrsim 6$ quasars on a logarithmic scale. 
While a substantial fraction of the sample quasars only have upper limits on $L_{\rm FIR}$, 
there seems to be a marginal correlation between the two quantities. 
Fitting a linear regression, including the upper limits, we find 
\begin{equation}\label{eq5}
\log \left( \frac{L_{\rm FIR}}{L_\odot} \right) = (2.94 \pm 1.47) + (0.69 \pm 0.11) \log \left( \frac{L_{\rm Bol}}{L_\odot} \right), 
\end{equation}
using the IRAF STSDAS package\footnote{STSDAS is a product of the Space Telescope Science Institute, which is operated by AURA for NASA.}. 
This relation is consistent with that found in \citet{2011AJ....142..101W} for stacked averages of quasars at $2 < z  < 7$. 
Note that this correlation is strengthened by adding low-luminosity objects like the HSC quasars, 
as the Spearman rank coefficient of the $\log L_{\rm Bol} - \log L_{\rm FIR}$ relation decreases from 0.51 (full sample in Figure \ref{fig9}) 
to 0.34 when restricting the sample to optically luminous ($L_{\rm Bol} > 10^{13}~L_\odot$) quasars only. 
This is qualitatively consistent with \citet{2018ApJ...866..159V}, 
who reported no significant correlation between these two quantities 
when focusing on optically luminous quasars ($L_{\rm Bol} \gtrsim 10^{13}~L_\odot$). 
The physical origin of this kind of correlation remains unclear, but one simple explanation 
is that both the quasar activity and star-forming activity are supported 
by a common source of gas, such as a $\sim 100$ pc scale circumnuclear gas disk 
seen in nearby Seyfert galaxies \citep[e.g.,][]{2016ApJ...827...81I}, 
as also found in some simulations \citep[e.g.,][]{2010MNRAS.407.1529H}. 
Note that, whatever the origin of the correlation is, time-variability 
of the quasar luminosity would weaken the correlation, 
given that star-formation occurs on a much longer time-scale 
than does nuclear mass accretion 
\citep{2011ApJ...737...26N,2014ApJ...782....9H}. 

In Figure \ref{fig9}, we can define a line of {\it parallel growth} \citep[e.g.,][]{2013ApJ...770...13W,2014A&A...566A..53D}. 
This indicates an evolutionary path in which SMBHs grow in tandem with their host galaxies 
to give rise to the local $M_{\rm BH} - M_{\rm bulge}$ relation. 
Here we adopt the local relation of \citet{2013ARA&A..51..511K}, 
i.e., $M_{\rm BH} \simeq 0.005 \times M_{\rm bulge}$. 
The parallel growth line is thus equivalent to $\dot{M}_{\rm BH} = 0.005 \times SFR$ (Figure \ref{fig9}). 
The line moves upward in the figure if we adopt the \citet{2013ApJ...764..184M} relation instead ($M_{\rm BH} \simeq 0.003 \times M_{\rm bulge}$). 
$L_{\rm Bol}$ and $\dot{M}_{\rm BH}$ are equated as $\dot{M}_{\rm BH} = ((1-\eta)/\eta) \cdot (L_{\rm Bol}/c^2)$, 
where $\eta$ is the radiative efficiency \citep[$\sim 0.1$,][]{2005ApJ...633..624V,2019arXiv190207056Z} and $c$ is the speed of light. 
It is evident that most of the $z \gtrsim 6$ quasars discussed here, including the low-luminosity objects, 
are on or below the parallel growth line (quasar-dominant phase) in this differential Magorrian plot. 

This result supports the evolutionary scenario that sequentially links vigorous starburst and SMBH growth 
\citep[e.g.,][]{1988ApJ...328L..35S,2006ApJS..163....1H,2008ApJS..175..356H,
2008AJ....135.1968A,2012NewAR..56...93A,2012MNRAS.426.3201S}. 
In this scenario, the peak of the quasar luminosity comes after the peak of the dust-obscured starburst phase, 
in which AGN feedback would expel the surrounding dusty ISM and quenches the starburst. 
As we saw, the low-luminosity quasars, including the HSC quasars, 
particularly hosted in massive galaxies ($M_{\rm dyn} \gtrsim 4 \times 10^{10}~M_\odot$) 
are located on or below the local Magorrian relation (Figure \ref{fig8}). 
Given the low $M_{\rm dust}$ of these low-luminosity quasars (mostly on the order of 10$^7$ $M_\odot$), 
an expected gas mass fraction in $M_{\rm dyn}$ after multiplying a gas-to-dust mass ratio \citep[e.g., 100,][]{2007ApJ...663..866D} is small, 
indicating that stellar masses dominate their $M_{\rm dyn}$. 
Meanwhile, the on-going star formation in their hosts are also weak; 
they are entering a quiescent phase \citep{2018PASJ...70...36I}. 
Note that this view is sensitive to the assumed $T_{\rm d}$ to compute 
$L_{\rm FIR}$ from the single ALMA photometry. 
We recall, however, that now we assume the same high $T_{\rm d}$ for the HSC quasars 
as the one canonically adopted for the optically and FIR much brighter quasars (= 47 K). 
It may be more plausible for the HSC quasars to have lower $T_{\rm d}$ such as 35 K (see also Section 3.2). 
If this is true, the HSC quasars will reside in a further quasar-dominant region on Figure \ref{fig9}. 
Future multi-band ALMA observations will enable to better constrain the $T_{\rm d}$ of these HSC quasars, 
as has been performed in some luminous quasars \citep[e.g.,][]{2019arXiv190606801W}. 
In summary, we suggest that these HSC quasars would have experienced an earlier vigorous starburst phase 
to generate their host galaxy masses, prior to the currently observed quasar-dominant phase, 
in order for them to reach the local Magorrian relation. 

Prime candidates for such starbursting predecessors are SMGs at even higher redshifts. 
So far, only three SMGs have been spectroscopically identified at $z > 6$: 
SPT0311$-$58 at $z = 6.90$ \citep{2018Natur.553...51M}, HFLS3 at $z = 6.34$ \citep{2013Natur.496..329R}, 
and HATLAS G09 83808 at $z = 6.03$ \citep{2018NatAs...2...56Z}.  
These are known to host extreme starbursts with SFR 
$\sim 3000$ $M_\odot$ yr$^{-1}$ (for the former two) and $\sim 380$ $M_\odot$ yr$^{-1}$ (for G09 83808, after correcting for gravitational magnification), 
which has allowed their host galaxies to grow as massive as $\sim 10^{11}~M_\odot$ within a fairly short time. 
This picture is consistent with recent simulation work of merger-induced galaxy evolution \citep{2019MNRAS.483.1256G}, 
which suggested that $z \sim 7$ hyper-luminous infrared galaxies (HyLIRGs) 
are indeed the ancestors of $z \sim 6$ luminous quasars. 

Whilst it is quite difficult to determine if there is an AGN in these gas-rich and dusty systems at $z > 6$ using a current instrument, 
studies in the nearby universe have shown that IR-luminous or gas-rich systems 
tend to possess AGNs with high Eddington ratio \citep[e.g.,][]{2005ApJ...625...78H,2012ApJ...750...92X,2018PASJ...70L...2I}. 
Such a high Eddington ratio (or even super-Eddington) phase may be essential 
for cosmic SMBH growth \citep[e.g.,][]{2004A&A...420L..23K,2008ApJ...676...33D}. 
Furthermore, recent ALMA observations, supplemented by deep {\it Chandra} 7 Ms survey data, 
have revealed a high fraction of AGNs in moderate luminosity SMGs 
(i.e., 90$^{+8}_{-19}$\% of ULIRG-class objects and 57$^{+23}_{-25}$\% of LIRG-class objects are AGNs) 
at $z \sim 1-3$ \citep{2018ApJ...853...24U}. 
About two-thirds of their sample (25 SMGs in total) 
are in the starburst-dominant phase in terms of the differential Magorrian relation \citep[see also][]{2013ApJ...778..179W}, 
which is qualitatively consistent with some earlier works \citep[e.g.,][]{2005ApJ...632..736A,2008AJ....135.1968A}. 
Our results, along with these previous findings, support the idea that 
SMBHs and their host galaxies do not actually co-evolve in a synchronized way. 
Rather, there seems to be an evolutionary path from the starburst-dominant phase to the quasar-dominant phase.

\section{Summary}\label{sec5}
As a part of the SHELLQs project, a large optical survey of low-luminosity quasars 
($M_{\rm 1450} \gtrsim -25$ mag) at $z \gtrsim 6$ with the Subaru Hyper Suprime-Cam (HSC), 
we performed Cycle 5 ALMA follow-up observations toward three HSC quasars, 
in addition to our previous Cycle 4 study of four quasars. 
We thus continue to grow the sample of hosts of low-luminosity, high redshift quasars, 
from the pioneering works of \citet{2013ApJ...770...13W,2015ApJ...801..123W,2017ApJ...850..108W}. 
The main findings of this paper strengthen our previous arguments in \citet{2018PASJ...70...36I}, 
and are summarized as follows. 

\begin{itemize}
\item[1.] We detected \cii line emission in all three target 
HSC quasars, with \cii luminosities between $(2.4 - 9.5) \times 10^8~L_\odot$. 
These are consistent with the Cycle 4 measurements for other HSC quasars, 
but are one order of magnitude smaller than those measured in optically luminous quasars. 
The \cii line width shows no clear dependence on the quasar luminosity. 
\item[2.] Within a common 1\arcsec.0 aperture, 
we detected underlying FIR continuum emission from two of the three target quasars. 
It is intriguing that one HSC quasar shows a ULIRG-like FIR luminosity ($L_{\rm FIR} \simeq 2 \times 10^{12}~L_\odot$), 
while another object was not detected even with our sensitive ALMA observations (3$\sigma$ limit of $L_{\rm FIR} < 9 \times 10^{10}~L_\odot$). 
There is a wide spread in $L_{\rm FIR}$ among the HSC quasars, even though their quasar luminosities in rest-UV are comparable. 
These $L_{\rm FIR}$ are again an order of magnitude smaller than those typically found in optically luminous quasars. 
\item[3.] The spatial extents of the (barely resolved) [C\,\emissiontype{II}]-emitting regions of these HSC quasars are $\sim 2-4$ kpc, 
roughly consistent with previous measurements of both optically luminous and low-luminosity quasars. 
Thus the difference in $L_{\rm FIR}$ between optically luminous quasars and our HSC quasars 
roughly translates to the difference in FIR luminosity density. 
\item[4.] The $L_{\rm \cii}/L_{\rm FIR}$ ratios of the HSC quasars 
(except for the ULIRG-class object J2239+0207) are consistent with local star-forming galaxies. 
However, the ratio is an order of magnitude smaller in some optically and FIR-luminous quasars from the literature. 
We suggest that the high FIR luminosity density of the observed region 
and/or existence of a high temperature dust-emitting region may be the physical origin of this \cii deficit. 
\item[5.] From the dynamical measurements based on the thin rotating disk assumption, along with the $M_{\rm BH}$ data, 
we found that most of the HSC quasars and similarly low-luminosity $z \gtrsim 6$ CFHQS quasars ($M_{\rm 1450} \gtrsim -25$ mag) 
tend to lie on or even below the local $M_{\rm BH} - M_{\rm bulge}$ relation 
particularly at the high $M_{\rm dyn}$ range ($\gtrsim 4 \times 10^{10}~M_\odot$). 
This would require a quite rapid mechanism of co-evolution, such as merger-induced galaxy evolution, 
to grow both the hosts and black holes, given the high redshifts ($z \gtrsim 6$) of our sample quasars. 
Optically luminous quasars ($M_{\rm 1450} \lesssim -25$ mag), on the other hand, 
host {\it overmassive} SMBHs with respect to the local relation, 
while the discrepancy becomes less evident at the massive-end of $M_{\rm dyn} \gtrsim 10^{11}~M_\odot$. 
However, we also implied our current limitation to fully probe down the low-$M_{\rm BH}$ range ($< 10^8~M_\odot$), 
which will be the subject of future sensitive NIR observations. 
\item[6.] All $z \gtrsim 6$ quasars compiled in this work are located 
in the {\it quasar-dominant region} of the $L_{\rm Bol} - L_{\rm FIR}$ plane. 
As low-luminosity quasars seem to be consistent with the local $M_{\rm BH} - M_{\rm bulge}$ relation at least at the high galaxy-mass range, 
and are transforming into a quiescent population in terms of both SMBH growth and stellar mass accumulation, 
there should have been a starburst-dominant phase for them to gain their large host dynamical masses ($\sim 10^{10-11}~M_\odot$), 
prior to the currently observed quasar phase. 
Submillimeter galaxies (SMGs) at even higher redshifts ($z \gtrsim 7$) would be 
the prime candidate for such a starbursting ancestor. 
\end{itemize}

In this work, we demonstrated the importance of investigating low-luminosity quasars 
to probe the nature of early co-evolution of black holes and host galaxies in an unbiased way. 
The trends of low-luminosity quasars shown above are indeed clearly different from those of optically-luminous quasars, 
although our conclusions are based on the small sample. 
As the number of low-luminosity HSC quasars at $z > 6$ is growing dramatically, 
we can statistically confirm the trends we found thus far with sensitive ALMA surveys.

\bigskip
\begin{ack}
We deeply appreciate the anonymous referee for his/her thorough reading and useful comments which greatly improved this paper. 
We also appreciate Toshihiro Kawaguchi for his intensive advice on this work. 
This paper makes use of the following ALMA data: ADS/JAO.ALMA 2017.1.00541.S. 
ALMA is a partnership of ESO (representing its member states), NSF (USA) and NINS (Japan), 
together with NRC (Canada), MOST and ASIAA (Taiwan), and KASI (Republic of Korea), in cooperation with the Republic of Chile. 
The Joint ALMA Observatory is operated by ESO, AUI/ NRAO and NAOJ. 

The Hyper Suprime-Cam (HSC) collaboration includes the astronomical 
communities of Japan and Taiwan, and Princeton University. 
The HSC instrumentation and software were developed by the National Astronomical Observatory of Japan (NAOJ), 
the Kavli Institute for the Physics and Mathematics of the Universe (Kavli IPMU), 
the University of Tokyo, the High Energy Accelerator Research Organization (KEK), 
the Academia Sinica Institute for Astronomy and Astrophysics in Taiwan (ASIAA), and Princeton University. 
Funding was contributed by the FIRST program from Japanese Cabinet Office, 
the Ministry of Education, Culture, Sports, Science and Technology (MEXT), 
the Japan Society for the Promotion of Science (JSPS), Japan Science and Technology Agency (JST), 
the Toray Science Foundation, NAOJ, Kavli IPMU, KEK, ASIAA, and Princeton University. 

This paper makes use of software developed for the Large Synoptic Survey Telescope. 
We thank the LSST Project for making their code available as free software at http://dm.lsstcorp.org. 

The Pan-STARRS1 Surveys (PS1) have been made possible 
through contributions of the Institute for Astronomy, 
the University of Hawaii, the Pan-STARRS Project Office, 
the Max-Planck Society and its participating institutes, 
the Max Planck Institute for Astronomy, Heidelberg and the Max Planck Institute for Extraterrestrial Physics, Garching, 
The Johns Hopkins University, Durham University, the University of Edinburgh, Queen's University Belfast, 
the Harvard-Smithsonian Center for Astrophysics, the Las Cumbres Observatory Global Telescope Network Incorporated, 
the National Central University of Taiwan, the Space Telescope Science Institute, 
the National Aeronautics and Space Administration under Grant No. NNX08AR22G 
issued through the Planetary Science Division of the NASA Science Mission Directorate, 
the National Science Foundation under Grant No. AST-1238877, 
the University of Maryland, and Eotvos Lorand University (ELTE). 

T.I., M.I., Y.M., N.K., T.N., K.K., and Y.T. are supported by Japan Society for the Promotion of Science (JSPS) 
KAKENHI grant numbers JP17K14247, JP15K05030, JP17H04830, JP15H03645, JP17H01114, and JP17H06130, respectively. 
Y.M. is also supported by the Mitsubishi Foundation grant No. 30140. 
T.G. acknowledges the support by the Ministry of Science and Technology of Taiwan through grant105-2112-M-007-003-MY3. 
T.I. is supported by the ALMA Japan Research Grant of NAOJ Chile Observatory, NAOJ-ALMA-217. 
\end{ack}

\end{document}